\documentclass[sigconf]{acmart}

\AtBeginDocument{%
  }

\setcopyright{acmlicensed}
\copyrightyear{2024}
\acmYear{2024}
\acmConference[CIKM '24]{Proceedings of the 33rd ACM International Conference on Information and Knowledge Management}{October 21--25, 2024}{Boise, ID, USA}
\acmBooktitle{Proceedings of the 33rd ACM International Conference on Information and Knowledge Management (CIKM '24), October 21--25, 2024, Boise, ID, USA}
\acmDOI{10.1145/3627673.3679546}
\acmISBN{979-8-4007-0436-9/24/10}

\usepackage{multirow}
\usepackage{subfig}
\usepackage{amsmath,bm}
\usepackage{setspace}
\usepackage{soul}
\usepackage{color, xcolor}
\usepackage{balance}

\begin{document}

\title{Predicting Scientific Impact Through Diffusion, Conformity, and Contribution Disentanglement}

\author{Zhikai Xue}
\affiliation{%
  \institution{SEM, East China Normal University}
  \city{Shanghai}
  \country{China}
  \postcode{200062}
}
\email{zkxue@stu.ecnu.edu.cn}

\author{Guoxiu He}
\authornote{Corresponding author.}
\affiliation{%
  \institution{SEM, East China Normal University}
  \city{Shanghai}
  \country{China}
  \postcode{200062}
}
\email{gxhe@fem.ecnu.edu.cn}

\author{Zhuoren Jiang}
\affiliation{%
  \institution{SPA, Zhejiang University}
  \city{Hangzhou}
  \country{China}
  \postcode{310058}
}
\email{jiangzhuoren@zju.edu.cn}

\author{Sichen Gu}
\affiliation{%
  \institution{SEM, East China Normal University}
  \city{Shanghai}
  \country{China}
  \postcode{200062}
}
\email{scgu@stu.ecnu.edu.cn}

\author{Yangyang Kang}
\affiliation{
  \institution{PI, Zhejiang University}
  \city{Hangzhou}
  \country{China}
}
\email{yangyangkang@zju.edu.cn}

\author{Star Zhao}
\affiliation{
  \institution{IBD, Fudan University}
  \city{Shanghai}
  \country{China}
  \postcode{200433}
}
\email{xzhao@fudan.edu.cn}

\author{Wei Lu}
\affiliation{
  \institution{SIM, Wuhan University}
  \city{Wuhan}
  \country{China}
  \postcode{430072}
}
\email{weilu@whu.edu.cn}

\renewcommand{\shortauthors}{Zhikai Xue et al.}

\begin{abstract}
The scientific impact of academic papers is influenced by intricate factors such as dynamic popularity and inherent contribution.
Existing models typically rely on static graphs for citation count estimation, failing to differentiate among its sources. In contrast, we propose distinguishing effects derived from various factors and predicting citation increments as estimated potential impacts within the dynamic context.
In this research, we introduce a novel model, \textbf{DPPDCC}, which \textbf{D}isentangles the \textbf{P}otential impacts of \textbf{P}apers into \textbf{D}iffusion, \textbf{C}onformity, and \textbf{C}ontribution values. It encodes temporal and structural features within dynamic heterogeneous graphs derived from the citation networks and applies various auxiliary tasks for disentanglement. By  emphasizing comparative and co-cited/citing information and aggregating snapshots evolutionarily, 
DPPDCC captures knowledge flow within the citation network. Afterwards, popularity is outlined by contrasting augmented graphs to extract the essence of citation diffusion and predicting citation accumulation bins for quantitative conformity modeling. Orthogonal constraints ensure distinct modeling of each perspective, preserving the contribution value. To gauge generalization across publication times and replicate the realistic dynamic context, we partition data based on specific time points and retain all samples without strict filtering. Extensive experiments on three datasets validate DPPDCC's superiority over baselines for papers published previously, freshly, and immediately, with further analyses confirming its robustness.
Our codes and supplementary materials can be found at \textcolor{blue}{\href{https://github.com/ECNU-Text-Computing/DPPDCC}{Github}}.
\end{abstract}

\ccsdesc[500]{Information systems~Data mining}

\keywords{Impact Prediction, Citation Network, Dynamic Heterogeneous Graph, Graph Neural Network, Contrastive Learning}

\maketitle

\section{Introduction}
\label{sec:intro}

In paper retrieval and recommendation, ranking papers based on their scientific impact can aid researchers in delving into intricate research efforts.
This is particularly significant considering the exponential annual growth in the number of published papers \cite{lo2020s2orc, chu2021slowed, xue2023dgcbert}. Given the inherent difficulties in quantifying the precise numerical value of scientific impact, citation count is regularly employed as a rough approximation \cite{evans2009open, sinatra2016quantifying,jiang2021hints,geng2022dgni,yang2023catehgn}. Nevertheless, the current ranking depends on the cumulative number of citations received to date, which merely represents the impact within the prior research environment. Therefore, it becomes imperative to cultivate a sophisticated model that can assess the potential impact of a research paper, placing significant emphasis on the future \cite{wang2013ctask}.

At a given point in time (\textit{e.g.}, at present), estimating the anticipated increment in citations can effectively serve as a better approximate measure of the potential impact, allowing for a fair comparison between long-published papers and newly-published ones.
Actually, the citations received by papers are not solely derived from the influence of their contributions but are also affected by a multitude of other factors \cite{bornmann2008citingbehavior, case2000citationbehavior}. 
Hence, rather than directly forecasting original citation counts, it is more effective to disentangle citation increments received from extrinsic factors and the actual contributing factor, making impact prediction more practical and interpretable.
Afterwards, we can identify exceptional works amidst the vast array of publications by discerning the potential contributing impact. In this work, we emphasize the popularity factor of a paper as a prominent extrinsic factor affecting the increase of citations. More precisely, when referring to the popularity factor of a paper, we are considering its citation diffusion patterns and the collective conformity degree within the citation network.

Regarding \textbf{citation diffusion}, we posit that influential papers assume an amplifying effect in shaping the information dissemination process within the citation network \cite{page1998pagerank,wu2019dindex}.
This leads to increased visibility of papers connected to them. In other words, papers cited by highly-cited papers will receive significantly more citations than others. The statistics of the datasets provide empirical support. For instance, in the field of Computer Science, the former papers will receive an average of 1.5475 citations (after applying a logarithmic transformation), while the latter ones will only receive an average of 0.8580 citations. 
Towards \textbf{collective conformity}, existing studies have found that some researchers may tend to cite papers associated with well-known entities, irrespective of their actual relevance or genuine contribution \cite{thornley2015citationauthority}.
This phenomenon is substantiated by the future disparity (2.1917 vs. 0.8009) in papers with different accumulated citations, which aligns with the concept of the Matthew effect in the academic domain \cite{allison1974citationmatthew}. 

\begin{figure}[]
  \centering
  \includegraphics[width=0.44\textwidth]{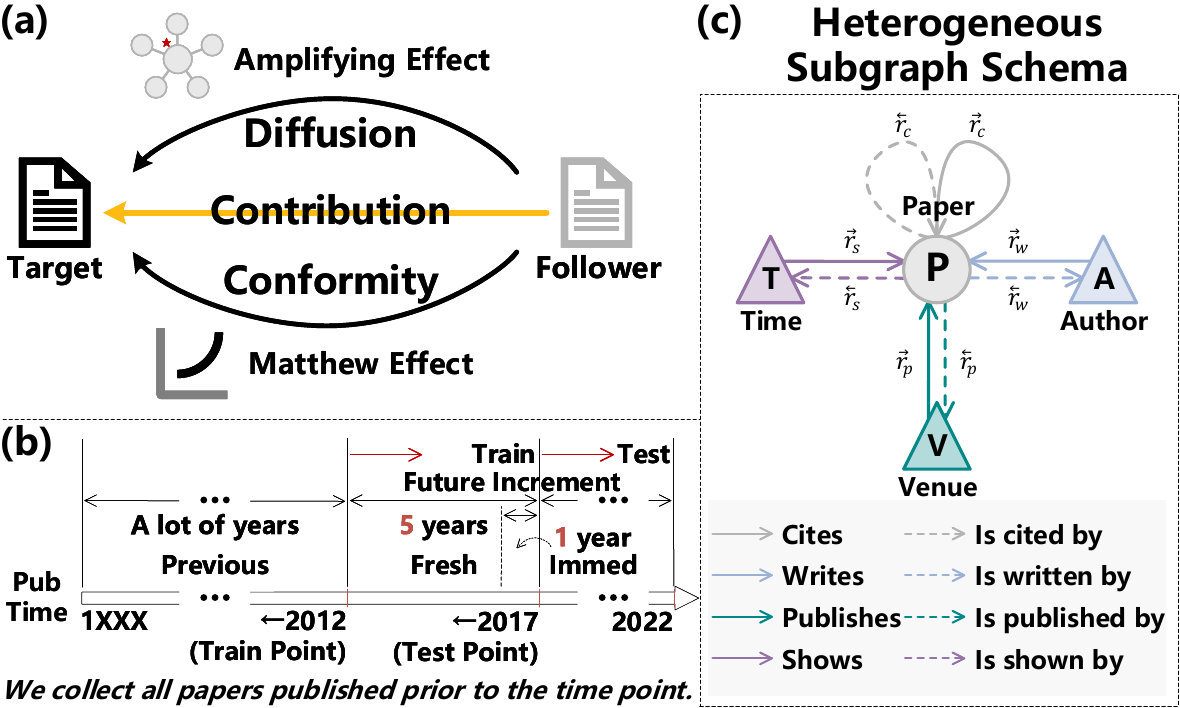}
    \caption{(a) We disentangle the citation increment into diffusion, conformity, and contribution values for better interpretability. (b) To imitate the evaluation of future dynamics within the citation context, we split the dataset by the observation time point and categorize samples into previous, fresh, and immediate papers in terms of publication time. (c) We represent the heterogeneous subgraph schema with all metadata, where each relation type is bidirectional.}
  \label{fig:intro}
\end{figure}

Therefore, by separating out the influence of the popularity, \textit{i.e.}, citation diffusion, and collective conformity, we can better approximate the genuine contributing impacts of papers. In essence, we aim to disentangle the citation increment into diffusion, conformity, and contribution values (as shown in Figure \ref{fig:intro}a). 

Typically, conventional research \cite{li2017deepcas, cao2017deephawkes, yang2022gtgcn, chen2022mucas, xu2022ccgl,he2023h2cgl} on citation prediction imposes strict constraints to select favorable samples by removing lowly-cited papers. 
Additionally, their division of data into training, validation, and testing sets is typically performed randomly, ignoring the chronological order of paper publications \cite{ruan2020predicting, abrishami2019predicting, huang2022fine, xue2023dgcbert}.
Moreover, some studies focus solely on the most recently published papers, neglecting the value of those published previously \cite{jiang2021hints, geng2022dgni, yang2023catehgn}.
These problem settings deviate from real-world scenarios, potentially causing a substantial distribution shift. Consequently, models may struggle to handle papers published at different times. 
Overall, the random data partitioning setting disregards new updates to the citation network, hampering the model's practicality and robustness in real-world applications, while the static problem setting impedes the exploration of the potential value of papers reactivated in the evolving context \cite{van2004sleeping}.

In contrast, our work emulates reality by revising the dynamic problem setting of previous studies \cite{he2023h2cgl}. We include all papers without strict filtering to maintain consistency with the real-world distribution of citation network data. As illustrated in Figure \ref{fig:intro}b, we partition the dataset based on specific observation time points, ensuring distinct training and testing windows. This methodology introduces \textbf{fresh} papers—excluded from the training set (\textbf{previous} papers) —as part of the test set, serving both as samples and context. We further consider \textbf{immediate} papers (published at the test time point) that closely align with cold-start scenarios, facilitating a more practical evaluation. Notably, previous papers have different contexts and future increments at the train and test time points. Adopting this strategy incentivizes our model to fit the dynamic citation context, thereby enhancing its generalization.

Existing graph disentanglement studies typically aim at implicitly decomposing abstractive hidden factors from either static or homogeneous graphs (see Section \ref{sec:relate}) \cite{ma2019disengcn, yang2020factorgnn, wang2020disenhan}. In contrast, our focus is explicitly disentangling citation behaviors' categorical characteristics within dynamic and heterogeneous citation graphs. For each target paper, we construct a Dynamic Heterogeneous Graph from the citation network, consisting of a sequence of heterogeneous snapshots (see Figure \ref{fig:intro}c), which retains both the content of the paper and its dynamic citation context.
Based on this graph, we devise a model named \textbf{DPPDCC} that \textbf{D}isentangles the \textbf{P}otential impacts of \textbf{P}apers into \textbf{D}iffusion, \textbf{C}onformity, and \textbf{C}ontribution values.
The model first employs a Citation-aware Graph Neural Network Encoder (CGE) to extract distinctive structural and temporal information alternately. The Popularity-aware Disentanglement Module (PDM) then separates the values derived from different factors through three simultaneous auxiliary tasks.
Specifically, the model utilizes Citation-aware Relational Graph Convolutional Networks (R-GCNs) to capture the structural features of the input snapshot sequence and a Transformer Encoder to incorporate temporal information. A novel message-passing module within the R-GCN, CompGAT, is designed to integrate comparative and co-cited/citing information between papers. Additionally, the Type-specific Attention Snapshot Readout is proposed to aggregate information from each snapshot evolutionarily.
Upon acquiring the final representation of the target paper, we disentangle the predicted increment into diffusion, conformity, and contribution values. For diffusion, we employ Triplet Citation-aware Graph Contrastive Learning to capture the popularity influence inherent in the information diffusion process within the citation network. For conformity, we introduce an auxiliary task to classify bins of accumulated citations. Furthermore, orthogonal constraints are imposed on the representations of disentangled perspectives in pairs, ensuring they encode distinct features without overlap.
Extensive experiments conducted on three fields of the real-world scientific dataset S2AG \cite{kinney2023s2orcnew} demonstrate DPPDCC's superior performance compared to baseline models across previous, fresh, and immediate papers. Ablation tests further highlight the significance of the proposed components, and visualizations provide insights into the model's interpretability.

To sum up, the contributions of this research are as follows: 

$\bullet$ We revise the problem formulation of impact prediction to better align with real-world scenarios. Without filtering predicted samples by their citation counts, we partition the datasets based on specific time points to maintain the dynamics of the citation context. Moreover, we employ disentangled representation learning to extract the contributing impacts of papers.

$\bullet$ We propose a novel model, DPPDCC, which first encodes comparative and co-cited/citing information evolutionarily within the Dynamic Heterogeneous Graph through a Citation-aware Graph Neural Network Encoder. It further disentangles the citation increment into diffusion, conformity, and contribution values.

$\bullet$ Experimental results show that DPPDCC notably outperforms existing baselines for previous, fresh, and immediate papers, and further analyses confirm our model's reasonable prediction ability.
\section{Related Work}
\label{sec:relate}

We review three lines of related work: Citation/Cascade Prediction, Dynamic and Heterogeneous Graph Neural Networks (GNNs), and Disentangled Representation Learning. 

Citation prediction is a vital sub-task of automatic academic evaluation as it enables the estimation of scientific impact. It is closely examined within cascade prediction, which shares analogous graph structures and objectives \cite{zhou2021cascade}.
Their approaches can be classified into three categories: stochastic, feature-based, and deep learning models.
Stochastic models predict future citation counts by fitting citation curves \cite{glanzel1995predictive}, following Zipf-Mandelbrot's law \cite{silagadze1997citations}. 
Recently, machine learning models have demonstrated promising results using manually extracted features from various metadata (\textit{e.g.} content, author, venue) \cite{yan2011citation,yu2014citation,ruan2020predicting,dong2015cikm}. More presently, deep neural networks have dominated by applying advanced Natural Language Processing (NLP) and Computer Vision (CV) methods to extract abstractive representations from paper content \cite{abrishami2019predicting,huang2022fine,cohan2020specter,xue2023dgcbert}.  
For cascade prediction, graph embeddings, sequence models, and GNN models extract structural information from the graph, with temporal information encoded by sequence models \cite{li2017deepcas,cao2017deephawkes,xu2022ccgl,cheng2023cassampling}.
However, most existing models exhibit suboptimal performance, failing to simultaneously exploit the valuable information presented in paper content and the scientific context within the citation network.
Furthermore, their strict data selection and static random data splitting strategies may hinder their practicality. 
In contrast, our study revises the problem setting of impact prediction in the dynamic context \cite{he2023h2cgl, yan2024cody}.
We ensure non-overlapping training and testing observation windows, retaining the real data distribution without sample filtering. We also utilize both paper content and dynamic context, and further introduce disentangled representation learning to provide a more practical impact estimation.

GNNs \cite{sperduti1997supervised,gori2005new,scarselli2008graph,wu2020comprehensive} are widely applied to handle non-Euclidean data like graphs. Dynamic graphs incorporate temporal information, while heterogeneous graphs involve multiple types of nodes or edges.
Dynamic graphs can be divided into snapshots at different time points. Previous models like DGCN \cite{manessi2020dgcn}, Dysat \cite{sankar2020dysat}, and ROLAND \cite{you2022roland} typically encode the snapshots with static GCN(s) as the structural encoder, followed by sequential models like RNN or Transformer Encoder. 
In heterogeneous graphs, R-GCN serves as a prominent model  \cite{schlichtkrull2018rgcn}. It conducts independent message passing within different relations and aggregates them to update node representations.
Our study proposes a Citation-aware GNN Encoder that effectively models both dynamic and heterogeneous graphs of target papers constructed from the citation network. It utilizes modified R-GCNs as structural encoders and a multi-layer transformer as the temporal encoder. Additionally, we incorporate bibliometric theory to model distinct features within the citation network.

Disentangled representation learning aims to identify and disentangle the underlying explanatory factors \cite{bengio2013replearning}. The existing efforts primarily focus on CV and NLP, utilizing Variational Auto-encoders \cite{ma2019disenrec,locatello2019challengingdisen,cai2019domaindisen,denton2017unsuperviseddisen,lee2021disenaug}. 
In graph learning tasks inspired by capsule networks, they partition node features into multiple hidden channels for implicit disentangled modeling \cite{ma2019disengcn}. Subsequent models strive to factorize the input graphs to facilitate distinct message passing \cite{yang2020factorgnn}. Recent models expand their scope to isolate causal and biased information presented in the graph structure and node features, with a particular emphasis on causal effects \cite{sui2022cal,fan2022disc}. In addition, certain studies extend the disentangled GNNs to specialized scenarios like recommendations and healthcare, addressing more complex graph structures involving either heterogeneous or dynamic graphs \cite{wang2020disenhan, zhang2022dida, wen2022ddhgnn,zhang2023sild,zhang2023dyted}.
In contrast to previous studies, our approach acknowledges that the potential impact within the citation network is influenced not only by the paper's contribution but also by popularity factors.
These popularity factors encompass the amplifying effect owing to the highly-cited paper nodes in information diffusion and the manifestation of the Matthew effect as collective conformity. 
Building upon previous disentangled studies in recommendations \cite{zheng2021dice, chen2022poprec}, we regard the final predicted values as mixed results of multiple factors and introduce auxiliary tasks for separate encoding of different perspectives. Leveraging the dynamic heterogeneous graph, we disentangle the potential impact into diffusion, conformity, and contribution values, aiming for a more rational estimation through explicit modeling.

\section{Methodology}
\label{sec:method}

\begin{figure*}[]
  \centering
  \includegraphics[width=0.80\textwidth]{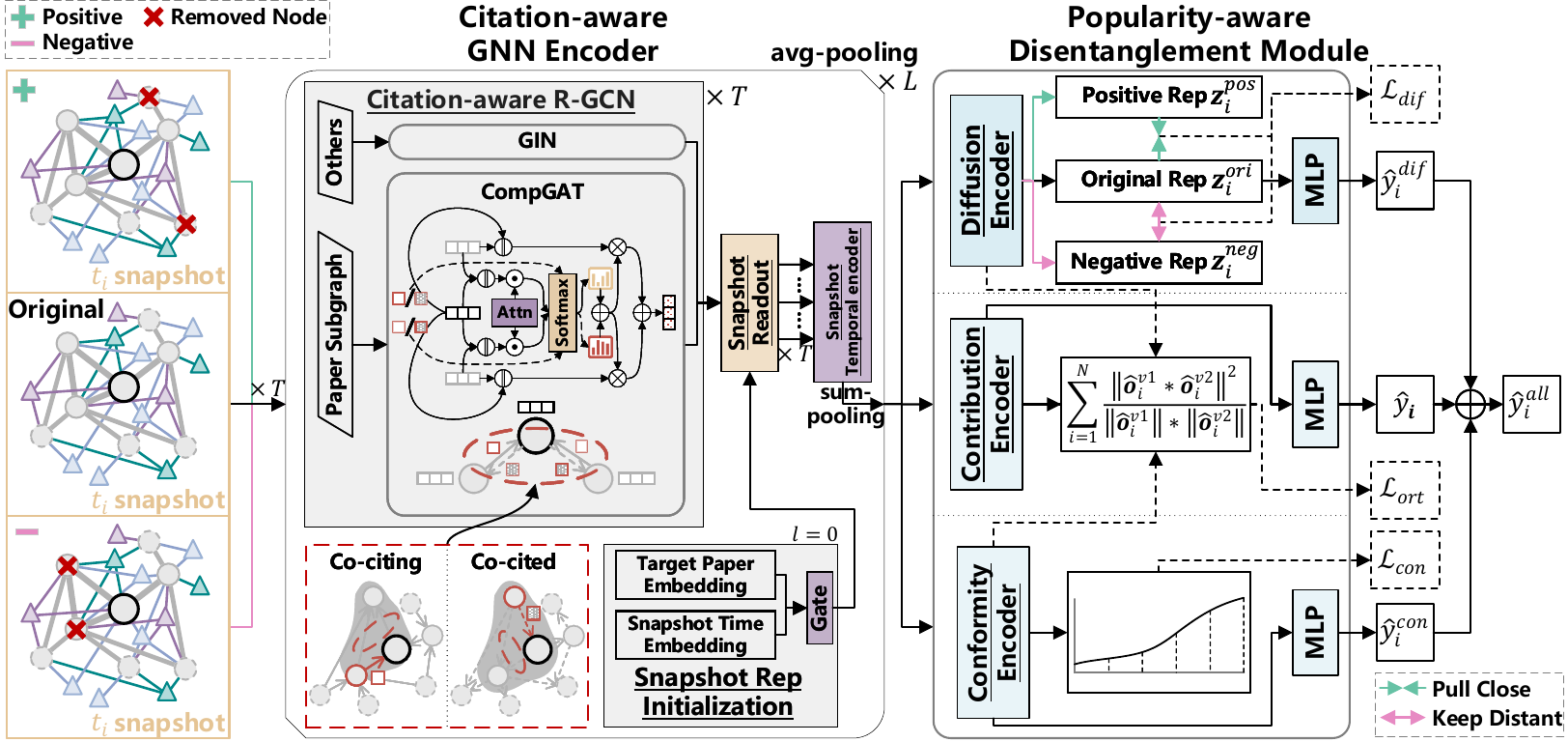}  
  \caption{The overall architecture of DPPDCC. DPPDCC first encodes the Dynamic Heterogeneous Graph of the target paper with the Citation-aware GNN Encoder. Based on the encoded representation of the target paper, DPPDCC then disentangles its citation increment into diffusion, conformity, and contribution values through three corresponding auxiliary tasks.}
  \label{fig:model}
\end{figure*}

In this section, we introduce our proposed model DPPDCC in detail. The overall framework of DPPDCC is depicted in Figure \ref{fig:model}, which can be divided into two parts: Citation-aware GNN Encoder (CGE) and Popularity-aware Disentanglement Module (PDM).

\subsection{Problem Formulation and Notations}
In practice, the citation network comprises diverse entities and relations that evolve dynamically each year. To capture the entity-aware and time-variant characteristics, we construct a dynamic heterogeneous graph from the citation network for each target paper to predict its potential impact.

\subsubsection{Dynamic Heterogeneous Graph}

Given a target paper at the time step $t$, its heterogeneous subgraph is defined as $G^t = (\nu^t, \epsilon^t, \phi, \varphi)$. 
Here, $\nu^t$ denotes the set of nodes surrounding the target paper at the time step $t$. And $\epsilon^t$ represents the set of edges among $\nu^t$. Each node $\nu$ and edge $\epsilon$ are associated with their type mapping functions $\phi:\nu \rightarrow U$ and $\varphi:\epsilon \rightarrow R$, where $U$ and $R$ denote the types of nodes and edges shared across all times (see Figure \ref{fig:intro}c), respectively.

The backbone of the heterogeneous subgraph primarily consists of the \textit{cited} and \textit{citing} relations. We collect $k$-hop cited/citing neighbors of the target paper at the specific time $t$. For each node in $i$-th hop, to maintain the most recent and significant information, we select the $K_i$ newest published and then most highly cited papers as neighbors. To address the potential lack of global information, we incorporate metadata nodes associated with the papers (see Figure \ref{fig:intro}c). Prior to the observation time point, we consider a total of $T$ such subgraphs, each serving as a snapshot captured at distinct time steps. These subgraphs organized into a sequence collectively constitute the dynamic heterogeneous graph $G$ of the target paper.

\subsubsection{Potential Impact Prediction}
We estimate the potential impact of the target paper by predicting its citation increment, defined as the increase in citation count $\Delta$ years after the observation time point. By disentangling the sources of this increment, our model can more accurately assess the genuine potential impact, thereby providing a closer approximation to the real scientific significance.

\subsection{Initialization}
The initial embedding of the \textit{paper} node is derived from the text embedding obtained by encoding the combination of title and abstract. Metadata node embeddings in each snapshot are the average embeddings over all papers associated with them in the global citation network at the corresponding time step.
Specifically, the extraction of text embeddings is accomplished using Sentence-BERT \cite{reimers2019sbert}.
To initialize the $0$-th layer representation of the snapshot of the $i$-th target paper at the $t$-th time step $\bm{h}_{ti, s}^{(0)}$, we mainly take into account the contextual information from the corresponding time embedding $\bm{h}_{t, time}^{(0)}$ of time step $t$. 
Since the time embedding only preserves the global features, we utilize a gate module to approach specific information from the $i$-th target paper embedding $\bm{h}_{i, p}^{(0)}$. The snapshot representation $\bm{h}_{ti, s}^{(l)}$ of $l$-th layer will then be the proxy to aggregate the corresponding heterogeneous subgraph:
\begin{equation}
    \begin{split}
        & \bm{h}_{ti, s}^{(0)} = \bm{a}_{ti} \times \bm{h}_{t, time}^{(0)} \\
        & \bm{a}_{ti} = \mathrm{sigmoid}(\bm{W}^{input}\bm{h}_{t, time}^{(0)} + \bm{W}^{target}\bm{h}_{i, p}^{(0)})
    \end{split}
\label{eq:snapshot_embs}
\end{equation}

\subsection{Citation-aware GNN Encoder}
Citation-aware GNN Encoder (CGE) aims to capture distinct structural and temporal features from the dynamic heterogeneous graph. To be specific, the backbone of CGE consists of modified R-GCNs \cite{schlichtkrull2018rgcn} as the structural encoder and a Transformer Encoder \cite{vaswani2017attention} as the temporal encoder. We stack $L$ integrated layers to encode them alternately instead of separate modeling. This allows us to exploit higher-order interactions between space and time.
To elaborate, for subgraphs at different time points, we first utilize distinct Citation-aware R-GCNs to model their structural information respectively. Afterwards, the Type-specific Attention Snapshot Readout is employed to summarize the information from these subgraphs based on the snapshot representations. These aggregated representations are then fed into the temporal encoder to capture the intricate temporal relationships between different time points. 
By integrating structural and temporal encoding, our approach could extract valuable insights from the dynamic heterogeneous graph.

\subsubsection{Citation-aware R-GCNs}
Citation-aware R-GCNs are utilized to encode node hidden states $\bm{h}$ within the dynamic heterogeneous graph for structural modeling:
\begin{equation}
    \begin{split}
        & \bm{h}_{dst}^{(l+1)} = \underset{r\in\mathcal{R}, r_{dst}=dst}{AGG} 
        (f_r(g_r, \bm{h}_{r_{src}}^{(l)}, \bm{h}_{r_{dst}}^{(l)})) \\
    \end{split}
\label{eq:graph_encoder}
\end{equation} 
where $\mathcal{R}$ is the set of all edges in the graph, $f_r$ is the message passing method applied to relation $r$, $g_r$ is the subgraph with relation $r$, $\bm{h}_{r_{src/dst}}$ is the hidden representation of the source/target node of relation $r$, and $AGG$ is the relation aggregation function.

Motivated by the Disruption-index \cite{wu2019dindex}, in "is cited by" edges, if the source and target paper have more identical references, the source paper may be less influenced by the target paper, resulting in a negative effect. Conversely, in "cites" edges, if the source and target paper have more identical citations, they may share more commonality and are more closely connected, leading to a positive effect. Thus, we introduce CompGAT, a novel module based on GATv2 \cite{brody2021gatv2}, to encode comparative and co-cited/citing features between \textit{paper} nodes. To capture the discrepancy between source and target nodes, we encode the concatenation of their representations. Besides the basic attention scores of GATv2, we incorporate co-cited or co-citing strengths by assessing the normalized similarity of their references or citations in the global citation network. Based on the adjacency matrix $\bm{A}$, it is calculated as $\bm{C} = \mathrm{norm}(\bm{A}\bm{A}^T)$, which represents 2-hop relationships connected with intermediate \textit{paper} nodes. Finally, the attention is a mixed distribution resembling CopyNet \cite{gu2016copynet}.
For $i$-th \textit{paper} node in $(l+1)$-th layer $\bm{h}_i^{(l+1)}$:
\begin{equation}
    \begin{split}
        & \bm{h}_i^{(l+1)} = \mathrm{LeakyReLU}\left(\mathrm{layernorm}(\bm{h}_i^{(l)} + \bm{h}_{in}^{(l+1)})\right) \\
        & \bm{h}_{in}^{(l+1)} = \sum_{j\in \mathcal{N}(i)} \alpha_{ij}^{(l)} \bm{W}_c^{(l)} [\bm{W}^{(l)}_{left} \bm{h}_{i}^{(l)} \| \bm{W}^{(l)}_{right} \bm{h}_j^{(l)}] \\
        & \alpha_{ij}^{(l)} = \lambda \mathrm{softmax_i^c} (c_{ij}) + (1 - \lambda)  \mathrm{softmax_i^e} (e_{ij}^{(l)})\\
        & e_{ij}^{(l)} = \bm{w}_a^{(l)}\mathrm{LeakyReLU}\left(\bm{W}^{(l)}_{left} \bm{h}_{i}^{(l)} + \bm{W}^{(l)}_{right} \bm{h}_{j}^{(l)}\right)
    \end{split}
\end{equation}
where $c_{ij} \in \bm{C}$ is the normalized co-cited/citing strength, $\lambda$ is to balance the sum distribution of co-cited/citing strengths and basic attention weights, and $\bm{w}_a^{(l)}$is the attention vector. Additionally, we apply GIN \cite{xu2018gin} with sum-pooling for other edges in the R-GCN.

\subsubsection{Type-specific Attention Snapshot Readout}
To represent the time-aware information of a single snapshot $\bm{h}_{t, s}$, we design an attention-based readout mechanism. It extracts the evolved temporal information of different types of \textit{paper} nodes (reference, citation, and target paper) based on the publication age of the target paper. 
\begin{equation}
    \begin{split}
        & \bm{h}_{t, s}^{(l+1)} = \bm{h}_{t, s}^{(l)} + \sum_{j\in \mathcal{N}(t)} \alpha_{tj}^{(l)} \bm{W}^{(l)}_{p} \bm{h}_{j, p}^{(l)} \\
        & \alpha_{tj}^{(l)} = \mathrm{softmax_t} (\hat{e}_{tj}^{(l)})\\
        & \hat{e}_{tj}^{(l)} = {\bm{w}_a}^{(l)}\mathrm{LeakyReLU}\left(\bm{W}^{(l)}_{s} \bm{h}_{t, s}^{(l)} + \bm{W}^{(l)}_{p} \bm{h}_{j, p}^{(l)}\right) + e_{tj}^{(l)} \\
        & e_{tj}^{(l)} = {\bm{w}_{t}}^{(l)}\mathrm{LeakyReLU}\left(f^{(l)}_{se}(\omega^{s}_{t}) + f^{(l)}_{pe}(\omega^{p}_{j})\right)
    \end{split}
\label{eq:readout}
\end{equation} 
where $\mathcal{N}(t)$ denotes the \textit{paper} nodes within $t$-th 1-hop subgraph, $f^{(l)}_{se}(\omega^{s}_{t})$ and $f^{(l)}_{pe}(\omega^{p}_{j})$ are snapshot and paper type embeddings. The snapshot type is determined by binning the time interval between time step $t$ and the publication time of the target paper.

\subsubsection{Snapshot Temporal Encoder} 
To encode the temporal information within snapshot representations across various times $\bm{H}_{s}^{(l+1)}$, we utilize a multi-layer Transformer Encoder \cite{vaswani2017attention}. 
Compared with classic sequential models, the self-attention mechanism captures more intricate relationships beyond strict ordinal dependencies: 
\begin{equation}
    \begin{split}
        & \hat{\bm{H}}_{s}^{(l+1)} = \mathrm{MultiHeadSelfAttn}(\bm{H}_{s}^{(l+1)})
    \end{split}
\end{equation}

\subsubsection{Final Output}
We obtain the target representation $\bm{o}$ by avg-pooling across the layers and then sum-pooling across the time steps to maintain high-order temporal information:
\begin{equation}
    \begin{split}
        & \Tilde{\hat{\bm{H}}}_{s} = avgpool^{L}([\hat{\bm{H}}_{s}^{(1)}, \hat{\bm{H}}_{s}^{(2)}, \cdots, \hat{\bm{H}}_{s}^{(L)}])\\
        & \bm{o} = sumpool^{T}(\Tilde{\hat{\bm{H}}}_{s})
    \end{split}
\label{eq:gnn_output}
\end{equation} 

\subsection{Popularity-aware Disentanglement Module}
We design the Popularity-aware Disentanglement Module (PDM) to disentangle the citation count potentially received from different perspectives, including citation diffusion, collective conformity, and actual contribution. 
Overall, we train the three components simultaneously and sum up their predicted values to obtain the original increment. We further apply different auxiliary tasks to help the disentanglement. This approach enables the model to learn distinctive compositions for diverse papers, thereby facilitating the identification of their authentic contributions.

Specifically, we utilize separate Multilayer Perceptrons (MLPs) as encoders and predictors to handle information within the target perspectives. For the target perspective $v$, we begin by deriving its representation $\hat{\bm{o}}_i^{v}$ from the original target representation $\bm{o}_i$ using its encoder $mlp^{v}_{enc}$. Subsequently, we input this representation into its corresponding predictor $mlp^{v}_{pred}$ to anticipate its increment $\hat{y}_i^{v}$:
\begin{equation}
    \begin{split}
        & \hat{\bm{o}}_i^{v} = mlp^{v}_{enc}(\bm{o}_i)\\
        & \hat{y}_i^{v} = mlp^{v}_{pred}(\hat{\bm{o}}_i^{v})\\
    \end{split}
\label{eq:predicted}
\end{equation}

\subsubsection{Diffusion Encoder}
The diffusion encoder captures the influence of popularity based on information diffusion, which is characterized by diversity affected by the degrees of spreading nodes. To extract the trunk of spreading, we propose a Citation-aware Triplet Graph Contrastive Learning method. We employ citation/degree-aware node-dropping \cite{xu2022ccgl, he2023h2cgl} as the graph augmentation technique.
In different snapshots, apart from the target \textit{paper} node, all other \textit{paper} nodes are selectively omitted with probability, which is weighted by their normalized global citations within the local subgraph. Lowly-cited \textit{paper} nodes are excluded for the positive view $z^{pos}$ since they contribute less to spreading. Conversely, for the negative view $z^{neg}$, highly-cited \textit{paper} nodes are omitted, given their significant impact on the graph semantics. This approach amplifies the influence of diffusion in the graph structure.
Contrasted with original sample $z^{ori}$, the diffusion loss $\mathcal{L}_{dif}$ is:
\begin{equation}
    \begin{split}
        & \bm{z}_i^{ori \vee pos \vee neg} = norm(mlp(\hat{\bm{o}}_i^{dif, ori \vee pos \vee neg}))\\
        & \mathcal{L}_{dif} = \frac {1}{N}\sum_{i=1}^{N}{-log\frac{exp(sim(\bm{z}^{ori}, \bm{z}^{pos})/\tau)}{\sum^{2}{exp(sim(\bm{z}^{ori}, \bm{z}^{neg \vee pos})/\tau)}}}
    \end{split}
\label{eq:dif_loss}
\end{equation}
where $sim(\cdot)$ is the similarity function like dot product and $\tau$ is the temperature parameter.

\subsubsection{Conformity Encoder}
The conformity encoder extracts the popularity attributed to the reputation of associated entities. Commonly, researchers tend to cite papers with significant accumulated citations instead of other potentially more relevant papers. In our work, we leverage the accumulated citations of the target paper at the predicted time as a signal of conformity. To extract its influence, we divide the predicted samples into equal frequency bins to quantify the Matthew effect. 
By predicting the binning label $y_{im}^{con}$, we prompt the model to learn the group differences resulting from the accumulated citations. Here, the conformity loss $\mathcal{L}_{con}$ is:
\begin{equation}
    \begin{split}
        & \hat{y}_{im}^{con} = mlp(\hat{\bm{o}}_i^{con})\\
        & \mathcal{L}_{con} = \frac{1}{N}\sum_{i=1}^{N} \mathcal{L}_i = - \frac{1}{N}\sum_{i=1}^{N} \sum_{m=1}^My_{im}^{con}\log(\hat{y}_{im}^{con}) 
    \end{split}
\label{eq:con_loss}
\end{equation}

\subsubsection{Contribution Encoder}
After disentangling the popularity influences, the remaining citation increment can serve as the genuine contribution.
To further encourage distinct encoding of various aspects for better approximation of its contribution, we apply the orthogonal regularization $\mathcal{L}_{ort}$ for each pair of perspectives:
\begin{equation}
    \begin{split}
        & \mathcal{L}_{ort} = \frac {1}{N}\sum_{i=1}^{N} \frac{\| \hat{\bm{o}}_i^{v1}*\hat{\bm{o}}_i^{v2} \|^2}{\|\hat{\bm{o}}_i^{v1} \|*\|\hat{\bm{o}}_i^{v2} \|}
    \end{split}
\end{equation}
where $\| \cdot \|$ is L2-norm, $v1$ and $v2$ are each pair of three perspectives.

\subsubsection{Disentanglement Loss}
We can obtain the final disentanglement loss $\mathcal{L}_{dis}$ by gathering the losses of different perspectives:
\begin{equation}
    \begin{split}
        & \mathcal{L}_{dis} =  \mathcal{L}_{dif} + \mathcal{L}_{con} + \mathcal{L}_{ort}
    \end{split}
\label{eq:dis_loss}    
\end{equation}

\subsection{Training}
Our task is to forecast the future increase in citations for the $i$-th paper. Considering that the distribution of citation count is skew distribution, we apply logarithmization to normalize the distribution. Thus, the primary loss function $\mathcal{L}_{reg}$ of the potential impact prediction task is the mean squared error (MSE) between logarithmized target and predicted citation increments: $y_i$ and $\hat{y}_i^{all}$.
\begin{equation}
    \begin{split}
        & \hat{y}_i^{all} = \hat{y}_i^{dif} + \hat{y}_i^{con} + \hat{y}_i\\
        & \mathcal{L}_{reg} = \frac {1}{N} \sum_{i=1}^{N}{(y_i - \hat{y}_i^{all})^2}
    \end{split}
\label{eq:reg_loss}
\end{equation}

The final loss function $\mathcal{L}$ includes the main regression loss and the disentangled learning loss:
\begin{equation}
    \begin{split}
        & \mathcal{L} = \mathcal{L}_{reg} + \beta  \mathcal{L}_{dis}
    \end{split}
\label{eq:all_loss}    
\end{equation}
where $\beta$ controls the strength of disentanglement tasks.

\section{Experiment}
\label{sec:exp}
In this section, we conduct extensive experiments and analyses on three subsets extracted from the real-world dataset S2AG to validate DPPDCC, aiming to answer the following research questions (RQs):

$\bullet$ \textbf{RQ1:} Can DPPDCC improve the impact prediction task?

$\bullet$ \textbf{RQ2:} What is the role of each component in DPPDCC? 

$\bullet$ \textbf{RQ3:} How sensitive is DPPDCC to hyper-parameters?

$\bullet$ \textbf{RQ4:} How does DPPDCC conduct the prediction?

\subsection{Experimental Settings}
\subsubsection{Datasets}
\begin{table}[htbp]
\setstretch{0.6}
  \centering
  \caption{Statistics for nodes and edges of the last citation network, as well as sample information of all sets.}
    \begin{tabular}{c|l|r|r|r}
    \toprule
    \multicolumn{2}{c|}{dataset} & \multicolumn{1}{c|}{CS} & \multicolumn{1}{c|}{CHM} & \multicolumn{1}{c}{PSY} \\
    \midrule
    \multirow{4}[2]{*}{node} & paper & 1,628,853  & 1,376,599  & 1,297,771  \\
          & author & 1,598,925  & 1,946,073  & 1,585,595  \\
          & venue & 12,524  & 8,389  & 13,775  \\
          & time  & 147   & 187   & 202  \\
    \midrule
    \multirow{4}[2]{*}{edge} & cite  & 11,534,431  & 10,382,698  & 13,401,112  \\
          & write & 5,123,460  & 6,355,630  & 4,813,135  \\
          & publish & 1,566,442  & 1,318,158  & 1,230,440  \\
          & have  & 1,628,853  & 1,376,599  & 1,297,771  \\
    \midrule
    \multirow{4}[4]{*}{sample} & train (2012) & 183,105  & 227,304  & 219,499  \\
          & val (2014) & 224,086  & 256,403  & 250,229  \\
          & test (2017) & 300,000  & 300,000  & 300,000  \\
\cmidrule{2-5}          & qualified pool & 1,112,611  & 1,093,660  & 1,015,579  \\
    \bottomrule
    \end{tabular}%
  \label{tab:data}%
\end{table}%

To establish the global citation network, we locate papers from S2AG \cite{kinney2023s2orcnew} that possess complete metadata or are high-impact despite no venue.
To select samples for testing, we target papers published prior to the test observation point (see Figure \ref{fig:intro}b). These papers should have complete metadata and contain at least one reference, without any restrictions on the citations they have received. 
We randomly sample 300,000 papers from the qualified pool to form the test set. 
Additionally, we gather all the metadata associated with these papers. 
To imitate the practical scenario, the training observation point is set 5 years before the test observation point (equal to the observation time window $T$ and predicted interval $\Delta$ for non-overlapping). The validation observation point precedes the test observation point by 3 years. 
Given the data quality and the disciplinary differences, we select three fields from the dataset: computer science (CS), chemistry (CHM), and psychology (PSY).
The relevant statistics are shown in Table \ref{tab:data}.

\begin{table*}[htbp]
\setstretch{0.6}
  \centering
  \caption{Performance comparison of DPPDCC with baselines in MALE and LogR$^2$. We divide the results into four categories: previous, fresh, and immediate papers, as well as the total. The best, second-best and third-best results are in \sethlcolor{red!20}\hl{red}, \sethlcolor{cyan!20}\hl{blue}, and \sethlcolor{green!20}\hl{green}, respectively. Significant improvements over best baseline results (with a 95\% confidence interval) are marked with $^*$.}
    \begin{tabular}{l|rrrr|rrrr|rrrr}
    \toprule
          \multicolumn{1}{c|}{Dataset} & \multicolumn{4}{c|}{Computer Science} & \multicolumn{4}{c|}{Chemsitry} & \multicolumn{4}{c}{Psychology} \\
    \midrule
    \midrule
    \multicolumn{1}{c|}{MALE($\downarrow$)} & \multicolumn{1}{c}{total} & \multicolumn{1}{c}{prev} & \multicolumn{1}{c}{fresh} & \multicolumn{1}{c|}{immed} & \multicolumn{1}{c}{total} & \multicolumn{1}{c}{prev} & \multicolumn{1}{c}{fresh} & \multicolumn{1}{c|}{immed} & \multicolumn{1}{c}{total} & \multicolumn{1}{c}{prev} & \multicolumn{1}{c}{fresh} & \multicolumn{1}{c}{immed} \\
    \midrule
    SciBERT & 0.6904  & 0.6110  & 0.8148  & 0.7996  & 0.5868  & 0.5384  & 0.7380  & 0.6841  & 0.6131  & 0.5776  & 0.7098  & 0.6828  \\
    SPECTER2 & 0.7482  & 0.6780  & 0.8583  & 0.8226  & 0.5832  & 0.5314  & 0.7450  & 0.6957  & 0.5935  & 0.5604  & 0.6837  & 0.6527  \\ 
    HINTS & 0.9022  & 0.9139  & 0.8838  & 0.9025  & 0.8265  & 0.7952  & 0.9246  & 0.9314  & 0.8529  & 0.8510  & 0.8582  & 0.8214  \\    
    CasSampling & \sethlcolor{green!20}\hl{0.5061}  & \sethlcolor{green!20}\hl{0.4249}  & \sethlcolor{green!20}\hl{0.6334}  & 0.8535  & \sethlcolor{green!20}\hl{0.4679}  & \sethlcolor{green!20}\hl{0.4266}  & \sethlcolor{green!20}\hl{0.5971}  & 0.8389  & \sethlcolor{green!20}\hl{0.5196}  & \sethlcolor{green!20}\hl{0.4757}  & 0.6392  & 0.8066  \\
    H$^2$CGL & \sethlcolor{cyan!20}\hl{0.4579}  & \sethlcolor{cyan!20}\hl{0.3870}  & \sethlcolor{cyan!20}\hl{0.5688}  & \sethlcolor{cyan!20}\hl{0.6455}  & \sethlcolor{cyan!20}\hl{0.4337}  & \sethlcolor{cyan!20}\hl{0.3982}  & \sethlcolor{cyan!20}\hl{0.5447}  & \sethlcolor{cyan!20}\hl{0.5820}  & \sethlcolor{cyan!20}\hl{0.4727}  & \sethlcolor{cyan!20}\hl{0.4415}  & \sethlcolor{cyan!20}\hl{0.5581}  & \sethlcolor{cyan!20}\hl{0.5848}  \\
    \midrule
    EGCN  & 0.8051  & 0.7865  & 0.8341  & 0.8322  & 0.6381  & 0.6135  & 0.7150  & 0.6894  & 0.7339  & 0.7307  & 0.7427  & 0.6978  \\
    Dysat & 0.6297  & 0.5571  & 0.7434  & 0.8129  & 0.5412  & 0.5036  & 0.6588  & 0.6774  & 0.6207  & 0.5874  & 0.7115  & 0.7138  \\
    ROLAND & 0.5978  & 0.5290  & 0.7056  & \sethlcolor{green!20}\hl{0.7384}  & 0.5437  & 0.4951  & 0.6955  & 0.6858  & 0.5521  & 0.5003  & 0.6934  & 0.7280  \\
    \midrule
    DisenGCN & 0.8321  & 0.7729  & 0.9249  & 0.9000  & 0.6343  & 0.6022  & 0.7347  & 0.6945  & 0.6689  & 0.6466  & 0.7296  & 0.7145  \\
    DisenHAN & 0.8221  & 0.7483  & 0.9379  & 1.0828  & 0.8652  & 0.9072  & 0.7341  & 0.6896  & 0.8766  & 0.8358  & 0.9877  & 0.9758  \\
    CAL   & 0.7094  & 0.6613  & 0.7848  & 0.7997  & 0.5733  & 0.5424  & 0.6698  & 0.6404  & 0.5535  & 0.5335  & \sethlcolor{green!20}\hl{0.6081}  & \sethlcolor{green!20}\hl{0.6324}  \\
    DisC  & 0.6733  & 0.6143  & 0.7656  & 0.7953  & 0.5609  & 0.5233  & 0.6784  & 0.6613  & 0.5592  & 0.5336  & 0.6289  & 0.6505  \\
    DIDA  & 0.6203  & 0.5403  & 0.7457  & 0.8096  & 0.5108  & 0.4707  & 0.6360  & 0.6757  & 0.5723  & 0.5389  & 0.6634  & 0.6765  \\
    DyTed & 0.6904  & 0.6110  & 0.8148  & 0.7996  & 0.5386  & 0.4960  & 0.6719  & 0.6784  & 0.5838  & 0.5451  & 0.6894  & 0.6986  \\
    SILD  & 0.5836  & 0.5061  & 0.7051  & 0.7738  & 0.5009  & 0.4679  & 0.6040  & \sethlcolor{green!20}\hl{0.6347}  & 0.5448  & 0.5122  & 0.6335  & 0.6566  \\
    \midrule
    DPPDCC & \sethlcolor{red!20}\hl{0.4432} & \sethlcolor{red!20}\hl{0.3724} & \sethlcolor{red!20}\hl{0.5540} & \sethlcolor{red!20}\hl{0.6232} & \sethlcolor{red!20}\hl{0.4264} & \sethlcolor{red!20}\hl{0.3949} & \sethlcolor{red!20}\hl{0.5250} & \sethlcolor{red!20}\hl{0.5688} & \sethlcolor{red!20}\hl{0.4587} & \sethlcolor{red!20}\hl{0.4287} & \sethlcolor{red!20}\hl{0.5405} & \sethlcolor{red!20}\hl{0.5699} \\
    \#improve (\%) & 3.21$^*$  & 3.78$^*$  & 2.60$^*$  & 3.45  & 1.69$^*$  & 0.84  & 3.62$^*$  & 2.27  & 2.96$^*$  & 2.88$^*$  & 3.14$^*$  & 2.55$^*$  \\
    \midrule
    \midrule
    \multicolumn{1}{c|}{LogR$^2$($\uparrow$)} & \multicolumn{1}{c}{total} & \multicolumn{1}{c}{prev} & \multicolumn{1}{c}{fresh} & \multicolumn{1}{c|}{immed} & \multicolumn{1}{c}{total} & \multicolumn{1}{c}{prev} & \multicolumn{1}{c}{fresh} & \multicolumn{1}{c|}{immed} & \multicolumn{1}{c}{total} & \multicolumn{1}{c}{prev} & \multicolumn{1}{c}{fresh} & \multicolumn{1}{c}{immed} \\
    \midrule
    SciBERT & 0.3370  & 0.4255  & 0.0859  & 0.0540  & 0.3887  & 0.3976  & 0.0209  & 0.0546  & 0.4761  & 0.5458  & 0.0779  & -0.0366  \\
    SPECTER2 & 0.2337  & 0.2944  & -0.0024  & 0.0125  & 0.3906  & 0.4101  & -0.0067  & 0.0120  & 0.5082  & 0.5714  & 0.1425  & 0.0449  \\
    HINTS & -0.1227  & -0.2590  & -0.1743  & -0.3801  & -0.1283  & -0.1891  & -0.5866  & -0.8466  & -0.0198  & -0.0021  & -0.3622  & -0.5247  \\
    CasSampling & \sethlcolor{green!20}\hl{0.6112}  & \sethlcolor{green!20}\hl{0.7002}  & \sethlcolor{green!20}\hl{0.4153}  & -0.1788  & \sethlcolor{green!20}\hl{0.6127}  & \sethlcolor{green!20}\hl{0.6331}  & 0.3371  & -0.4418  & \sethlcolor{green!20}\hl{0.6200}  & \sethlcolor{green!20}\hl{0.6945}  & 0.2431  & -0.4616  \\
    H$^2$CGL & \sethlcolor{cyan!20}\hl{0.6812}  & \sethlcolor{cyan!20}\hl{0.7268}  & \sethlcolor{cyan!20}\hl{0.5563}  & \sethlcolor{cyan!20}\hl{0.3846}  & \sethlcolor{cyan!20}\hl{0.6573}  & \sethlcolor{red!20}\hl{0.6582}  & \sethlcolor{cyan!20}\hl{0.4626}  & \sethlcolor{cyan!20}\hl{0.3064}  & \sethlcolor{cyan!20}\hl{0.6836}  & \sethlcolor{cyan!20}\hl{0.7315}  & \sethlcolor{cyan!20}\hl{0.4217}  & \sethlcolor{cyan!20}\hl{0.2209}  \\
    \midrule
    EGCN  & 0.1296  & 0.0704  & 0.0290  & -0.1268  & 0.2826  & 0.2080  & 0.0940  & 0.0350  & 0.2530  & 0.2654  & 0.0042  & -0.0896  \\
    Dysat & 0.4242  & 0.4812  & 0.2320  & -0.0057  & 0.4705  & 0.4499  & 0.2326  & 0.0784  & 0.4354  & 0.4881  & 0.0882  & -0.1431  \\
    ROLAND & 0.4911  & 0.5337  & 0.3314  & \sethlcolor{green!20}\hl{0.2044}  & 0.4495  & 0.4518  & 0.1344  & 0.0530  & 0.5718  & 0.6572  & 0.1419  & -0.1849  \\
    \midrule
    DisenGCN & 0.1026  & 0.1055  & -0.0849  & -0.1153  & 0.2897  & 0.2344  & 0.0502  & 0.0456  & 0.3746  & 0.4154  & 0.0549  & -0.1024  \\
    DisenHAN & -0.0488  & 0.0421  & -0.3820  & -0.8766  & -0.1110  & -0.3875  & 0.0577  & 0.0335  & -0.1186  & -0.0173  & -0.7948  & -1.1430  \\
    CAL   & 0.1859  & 0.2790  & -0.1024  & -0.2403  & 0.4166  & 0.3867  & 0.1753  & 0.1569  & 0.5783  & 0.6163  & \sethlcolor{green!20}\hl{0.3238}  & \sethlcolor{green!20}\hl{0.1045}  \\
    DisC  & 0.3496  & 0.3926  & 0.1605  & -0.0066  & 0.4259  & 0.4086  & 0.1538  & 0.0835  & 0.5606  & 0.6039  & 0.2820  & 0.0589  \\
    DIDA  & 0.4449  & 0.5228  & 0.2295  & 0.0043  & 0.5354  & 0.5318  & 0.2855  & 0.0847  & 0.5328  & 0.5887  & 0.2005  & -0.0299  \\
    DyTed & 0.3370  & 0.4255  & 0.0859  & 0.0540  & 0.4602  & 0.4454  & 0.2001  & 0.0821  & 0.5029  & 0.5654  & 0.1382  & -0.0967  \\
    SILD  & 0.4977  & 0.5809  & 0.2862  & 0.0503  & 0.5629  & 0.5539  & \sethlcolor{green!20}\hl{0.3440}  & \sethlcolor{green!20}\hl{0.1766}  & 0.5889  & 0.6444  & 0.2733  & 0.0260  \\
    \midrule
    DPPDCC & \sethlcolor{red!20}\hl{0.6938} & \sethlcolor{red!20}\hl{0.7349} & \sethlcolor{red!20}\hl{0.5775} & \sethlcolor{red!20}\hl{0.4236} & \sethlcolor{red!20}\hl{0.6667} & \sethlcolor{cyan!20}\hl{0.6577}  & \sethlcolor{red!20}\hl{0.5056} & \sethlcolor{red!20}\hl{0.3458} & \sethlcolor{red!20}\hl{0.6969} & \sethlcolor{red!20}\hl{0.7380} & \sethlcolor{red!20}\hl{0.4633} & \sethlcolor{red!20}\hl{0.2594} \\
    \#improve (\%) & 1.85$^*$  & 1.11$^*$  & 3.80$^*$  & 10.13  & 1.43$^*$  & \multicolumn{1}{c}{--}  & 9.29$^*$  & 12.84$^*$  & 1.94$^*$  & 0.89$^*$  & 9.87$^*$  & 17.44$^*$  \\
    \bottomrule
    \end{tabular}%
  \label{tab:comp}%
\end{table*}%

\subsubsection{Baseline models}
The baseline models are mainly divided into citation/cascade prediction models, dynamic GNN models, and disentangled GNN models:

\textbf{Citation/Cascade prediction models:}
We apply classic and recent citation/cascade prediction models to estimate the potential impacts of papers based on content, citation graphs, or citation cascade graphs. 
\textbf{SciBERT} \cite{beltagy2019scibert} is fine-tuned to encode the title and abstract of the paper.
\textbf{SPECTER2} \cite{cohan2020specter} focuses on document-level embeddings enhanced by a citation-aware pretrained task.
\textbf{HINTS} \cite{jiang2021hints} addresses the cold-start problem by applying GCN on pseudo meta-data graphs, using a stochastic model to make predictions.
\textbf{CasSampling} employs an importance-aware sampling technique to extract crucial segments of cascade graphs while preserving the temporal signals of the overall graphs.
\textbf{H$^2$CGL} \cite{he2023h2cgl} models citation network dynamics hierarchically and integrates contrastive learning for representation refinement. It shares a dynamic setup and a heterogeneous GNN framework similar to ours.

\textbf{Dynamic GNN models:}
We employ recent dynamic GNN models to extract paper representations from the dynamic graphs collected similar to our model, treating potential impact prediction as a graph regression task.
EvolveGCN \cite{pareja2020evolvegcn} integrates the RNN structure within GNN modules to facilitate sequential dynamic updates. \textbf{Dysat} \cite{sankar2020dysat} incorporates multiple self-attention modules to encode structural and temporal information separately. \textbf{ROLAND} \cite{you2022roland} is a novel dynamic heterogeneous GCN model that inherits static GCN methods hierarchically and homogeneously encodes heterogeneous information using type indicators. 

\textbf{Disentangled GNN models:}
We apply recent disentangled GNN models to extract critical representations from the multi-hop subgraphs. \textbf{DisenGCN} \cite{ma2019disengcn} leverages neighborhood routing from capsule networks for disentanglement. \textbf{DisenHAN} \cite{wang2020disenhan} builds upon DisenGCN and is adapted to heterogeneous graphs for the recommendation. \textbf{CAL} \cite{sui2022cal} employs attention modules to estimate the causal and trivial masks for structures and attributes and encodes them with separate GCNs. \textbf{DisC} \cite{fan2022disc} estimates the causal/bias masks and further employs causal/bias-aware loss functions as well as counterfactual unbiased training. \textbf{DIDA} \cite{zhang2022dida} discovers invariant patterns within dynamic graphs and applies random variant pattern swapping as the causal intervention. 
\textbf{DyTed} \cite{zhang2023dyted} devises a framework combining contrastive learning and adversarial learning to disentangle time-invariant and time-varying representations proficiently.
\textbf{SILD} \cite{zhang2023sild} expands upon DIDA in the spectral domain.

\subsubsection{Implementation details}
All models are implemented with PyTorch \cite{paszke2019pytorch} and DGL \cite{wang2019dgl}/PyG \cite{Fey/Lenssen/2019pyg}. 
Training is conducted on an NVIDIA A800 80GB GPU and optimized with the Adam optimizer \cite{kingma2014adam}. 
MALE (Mean Absolute Logarithmic Error) and LogR$^2$ (Logarithmic R-squared) are employed for evaluation.
We run the experiments with three random seeds and report the averaged results.
Models are developed using both training and validation sets, with the final model selected based on the best MALE on the validation set.
Each dataset is divided into previous, fresh, and immediate papers for detailed analysis. 
We adhere to the recommended settings in the original papers/codes for all baselines.
In our model, the hop order $k$ is set to 2, corresponding top limits $K_1$ and $K_2$ are set to 100 and 20, the number of layers $L$ is set to 4, and the co-cited/citing weight $\lambda$ and the disentanglement weight $\beta$ are set to 0.5.

\subsection{Performance Comparison (RQ1)}
To answer \textbf{RQ1}, we evaluate the performance of DPPDCC and other baselines. From Table \ref{tab:comp}, we have the following observations:

(1) \textbf{All models perform well on previous papers presented in the training set but experience degradation when dealing with fresh and immediate papers in the test set, particularly in terms of the LogR$^2$.} 
A substantial disparity is evident between the dynamic citation prediction models (H$^2$CGL and ours) and other models.
Even the best static model, CasSampling, struggles with immediate papers due to limited graph structures, tailored as it is a cascade prediction model designed for papers with sufficient citations. 
However, dynamic GNNs excel in encoding multi-hop information, compensating for the absence of cascade structures. 
Additionally, models like SciBERT, SPECTER2, and H$^2$CGL can leverage prior knowledge, such as content information or bibliometric theory, to mitigate these challenges.

(2) \textbf{Both dynamic and heterogeneous graphs are important for our task.} 
On one side, temporal information within dynamic graphs reveals the evolving trends, underscoring its critical role. DIDA, DyTed, and SILD, which consider dynamic graphs, stand out among all disentangled GNNs. On the other side, leveraging heterogeneous information can bolster the performance of dynamic GNNs, as demonstrated by ROLAND. However, proper modeling of heterogeneous information is essential; otherwise, it may result in performance degradation, as evidenced by DisenHAN compared to its backbone, DisenGCN.  
The top-performing baseline, H$^2$CGL, hierarchically encodes both dynamic and heterogeneous graphs, effectively adapting to the dynamic setting and achieving the best performance among baselines.

(3) \textbf{While disentanglement methods yield performance enhancements, their adaptation to dynamic heterogeneous graphs is crucial for achieving optimal flexibility.} Remarkably, SILD achieves superior outcomes compared to ROLAND in Chemistry, even when utilizing only homogeneous graphs. Other dynamic disentangled GNNs also showcase comparable performance, with DyTed outperforming its backbone, Dysat. Additionally, in the densest dataset, Psychology, disentangled GNNs demonstrate commendable performance, with CAL surpassing both CasSampling and ROLAND for fresh and immediate papers. 
However, in other scenarios, aside from dynamic disentangled GNNs, other methods struggle to compete with task-specific models and dynamic GNNs. 

(4) \textbf{Our proposed model surpasses all baselines by effectively modeling the dynamic heterogeneous graph and integrating task-adjusted disentanglement.}
This results in an average improvement of approximately 3\% in MALE across all categories, alongside a significant advancement of over 10\% in LogR$^2$ for fresh and immediate papers. 
Initially, compared to traditional citation/cascade prediction models, it demonstrates superior generalization and practicability through the fusion of content and context information within citation networks, coupled with its disentanglement capabilities.
Furthermore, we leverage the unique properties of citation networks to encode complex high-order structural and temporal information, thereby outperforming common dynamic GNNs.
Crucially, our model introduces interpretability from disentanglement models without sacrificing performance and further adapts to the task for enhanced improvement, thus surpassing the best dynamic citation prediction model, H$^2$CGL.

\subsection{Ablation Test (RQ2)}
To answer \textbf{RQ2}, ablation tests are performed on both CGE and PDM, highlighting the significance of various modules. For CGE, (-co-weight) removes the proposed co-cited/citing strengths, (-CompGAT) replaces the CompGAT with GATv2 \cite{brody2021gatv2}, and (-readout) simply obtains the snapshot representation through avg-pooling over all \textit{paper} nodes. For PDM, (-diffusion) removes the diffusion disentanglement, (-conformity) omits the conformity disentanglement, (-orthog) disregards the orthogonal constraints, and (-disen) eliminates all perspectives and replaces with a single MLP. 

\begin{table}[htbp]
\setstretch{0.7}
  \centering
    \caption{Results of ablation studies in all datasets.}
  \resizebox{0.47\textwidth}{!}{
    \begin{tabular}{l|rr|rr|rr}
    \toprule
    \multicolumn{1}{c|}{\multirow{2}[4]{*}{model}} & \multicolumn{2}{c|}{CS} & \multicolumn{2}{c|}{CHM} & \multicolumn{2}{c}{PSY} \\
\cmidrule{2-7}          & \multicolumn{1}{c}{MALE} & \multicolumn{1}{c|}{LogR$^2$} & \multicolumn{1}{c}{MALE} & \multicolumn{1}{c|}{LogR$^2$} & \multicolumn{1}{c}{MALE} & \multicolumn{1}{c}{LogR$^2$} \\
    \midrule
    \multicolumn{1}{c|}{complete} & \textbf{0.4432} & \textbf{0.6938} & \textbf{0.4264} & \textbf{0.6667} & \textbf{0.4587} & \textbf{0.6969} \\
    \midrule
    -co-weight & 0.4517  & 0.6855  & 0.4623  & 0.6106  & 0.4625  & 0.6953  \\
    -CompGAT & 0.4620  & 0.6842  & 0.4532  & 0.6202  & 0.4604  & 0.6921  \\
    -readout & 0.4982  & 0.6432  & 0.4404  & 0.6564  & 0.4867  & 0.6720  \\
    \midrule
    -diffusion & 0.4632  & 0.6801  & 0.4531  & 0.6231  & 0.4621  & 0.6914  \\
    -conformity & 0.4655  & 0.6814  & 0.4767  & 0.5799  & 0.4646  & 0.6935  \\
    -orthog & 0.4576  & 0.6870  & 0.4560  & 0.6168  & 0.4597  & 0.6958  \\
    -disen & 0.4597  & 0.6937  & 0.4696  & 0.5990  & 0.4830  & 0.6738  \\
    \bottomrule
    \end{tabular}%
  }
  \label{tab:ablation}%
\end{table}%

From Table \ref{tab:ablation}, we observe that:
(1) Since removing any module leads to a performance decrease, all modules play crucial roles in DPPDCC.
(2) For CGE, the removal of type-specific attention snapshot readout may cause a notable performance drop. 
That could be attributed to the presence of excessive noisy information within multi-hop snapshots, which necessitates the filtering of aggregated information for better concentration. 
CompGAT effectively captures distinct comparative information, and the incorporation of co-cited/citing strengths further complements the multi-hop information beyond the local neighborhood.
(3) Eliminating any disentanglement task adversely affects performance. Moreover, an evident decline is observed after removing the whole disentanglement module. 
Conformity rises as the paramount task among all disentanglement tasks. It focuses on the Matthew effect, which is widely pervasive and readily discernible in reality.

\subsection{Hype-parameters Test (RQ3)}

\begin{figure}[]
  \centering
  	\centering
 	\subfloat[$L$]{\includegraphics[height=0.088\textheight]{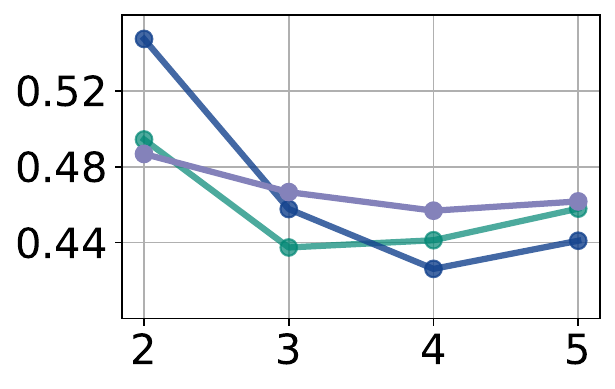}%
		\label{fig:param_layer}}
        \hfil
	\subfloat[$\lambda$]{\includegraphics[height=0.088\textheight]{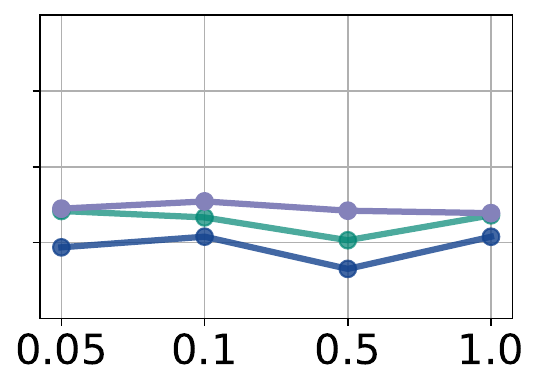}%
		\label{fig:param_lw}}
        \hfil
	\subfloat[$\beta$]{\includegraphics[height=0.088\textheight]{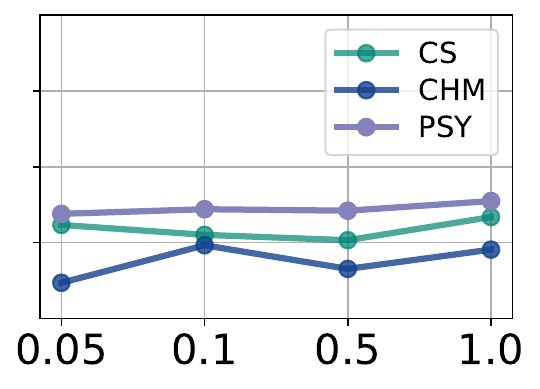}%
		\label{fig:param_dw}}
          \hfil
  \caption{Results of hyper-parameters test in terms of MALE.}
\label{fig:hyper}
\end{figure}

To address \textbf{RQ3}, we aim to explore the model's sensitivity to essential hyper-parameters, such as $L$, $\lambda$, and $\beta$, which respectively control the order/hop information of CGE, the importance of co-cited/citing strengths, and the weight of the disentanglement tasks relative to the main task. The results about MALE with varying settings are depicted in Figure \ref{fig:hyper}. We observe that: (1) When the layer number $L$ continues to increase, the performance will first increase and then decrease. Since we stack the structural and temporal encoders simultaneously, their interaction is more complicated and may also face the over-smoothing problem.
(2) The co-cited/citing weight $\lambda$ and disentanglement weight $\beta$ demonstrate greater stability compared to $L$, especially in Psychology. Selecting the proper hyper-parameters can further improve the performance. 
\subsection{Disentanglement Visualization (RQ4)}
\label{subsec:disen_visual}
\begin{figure}[]
  \centering
        \subfloat[time trend]{\includegraphics[height=0.088\textheight]{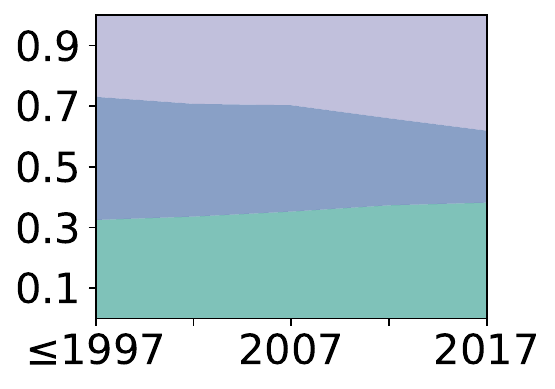}%
		\label{fig:vis_time}}
        \hfil
	\subfloat[group detail]{\includegraphics[height=0.088\textheight]{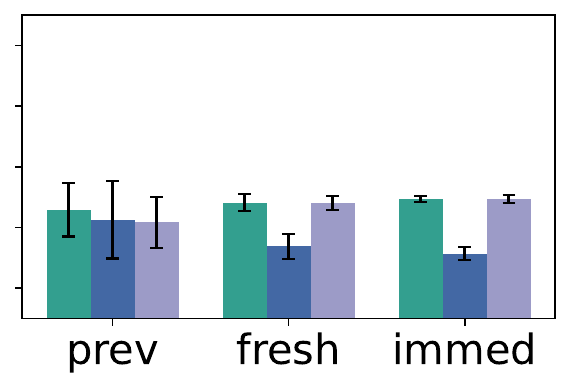}%
		\label{fig:vis_group}}
        \hfil
	\subfloat[value bin]{\includegraphics[height=0.088\textheight]{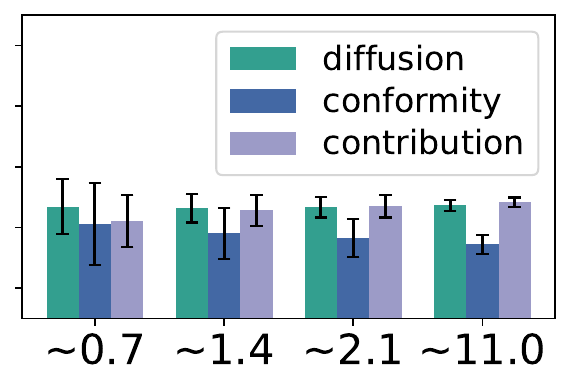}%
		\label{fig:vis_vbin}}
        \hfil
  \caption{Visualization of disentangled value proportions in CS. (a) displays the trend that evolved with the publication time. (b) demonstrates the detailed composition of papers categorized into previous, fresh, and immediate ones. (c) is binning the samples based on the predicted values.}
\label{fig:vis}
\end{figure}

To address \textbf{RQ4}, we provide visualizations of the disentanglement composition and the conformity representation distribution. 
Additional information can be found in the supplementary materials.
\subsubsection{Disentanglement Composition Visualization}
In Figure \ref{fig:vis}, we visualize samples with all positive values in the test set of Computer Science. 
Papers containing negative disentangled values typically exhibit nearly no increase in citations. Remarkably, these papers consistently feature negative values because their contribution values are the lowest across all perspectives, demonstrating our model's rationality.
In the stacked plot associated with time in Figure \ref{fig:vis_time}, a clear trend emerges: the contribution proportion gradually increases, while the conformity proportion undergoes a significant drop. This trend is logical, as older papers tend to accumulate substantial citations, thereby attracting more researchers to cite them due to inertia. Additionally, the diffusion proportion appears more stable over time, as it represents potential hidden properties within diffusion. This stability becomes more pronounced when papers are categorized into previous, fresh, and immediate ones, as shown in Figure \ref{fig:vis_group}. Diffusion proportions remain steady in both mean and standard deviation.
Moreover, contribution emerges as the most critical perspective, with its dominant and variable proportion. It holds particular significance for fresh and immediate papers. For immediate papers receiving almost no citations, both the diffusion and conformity perspectives struggle to offer predictive insights. However, the contribution can distinguish their potential in advance based on the citation context.
Furthermore, in Figure \ref{fig:vis_vbin}, when we bin the samples based on the predicted values, we observe a descending trend for conformity and a growing trend for contribution. Notably, conformity is on par with contribution within the lowest bin group, while diffusion accounts for the largest proportion. However, for papers with higher increments, contribution plays the most significant role and surpasses all other perspectives.

\subsubsection{Disentanglement Representation Visualization}
    \begin{figure}[]
      \centering
        \centering
        \subfloat[total]{\includegraphics[width=0.11\textwidth]{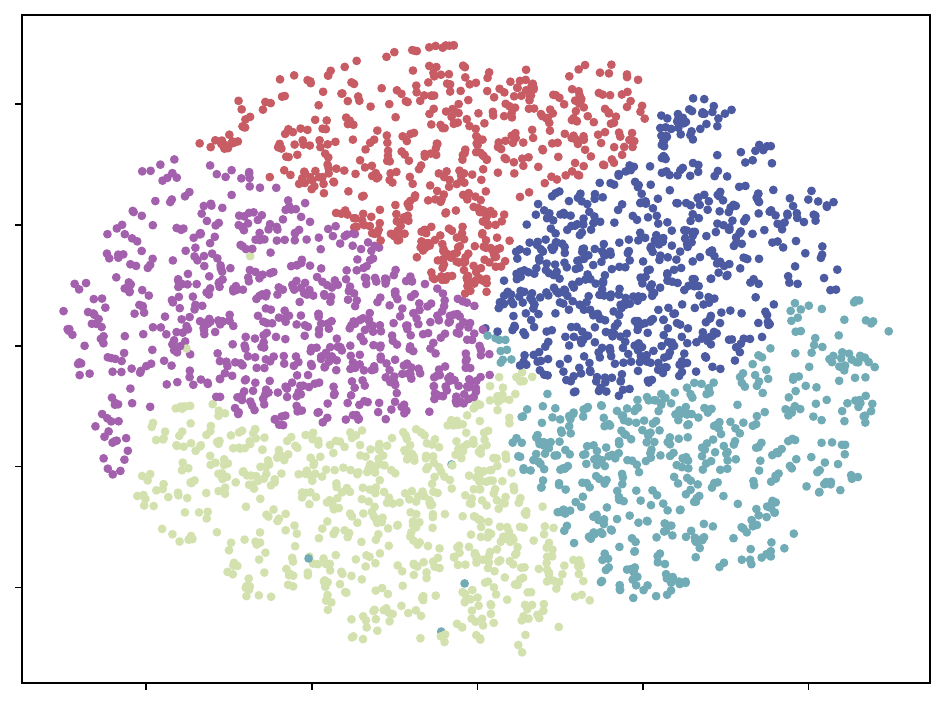}%
            \label{fig:rep_total_conf}}
            \hfil
        \subfloat[prev]{\includegraphics[width=0.11\textwidth]{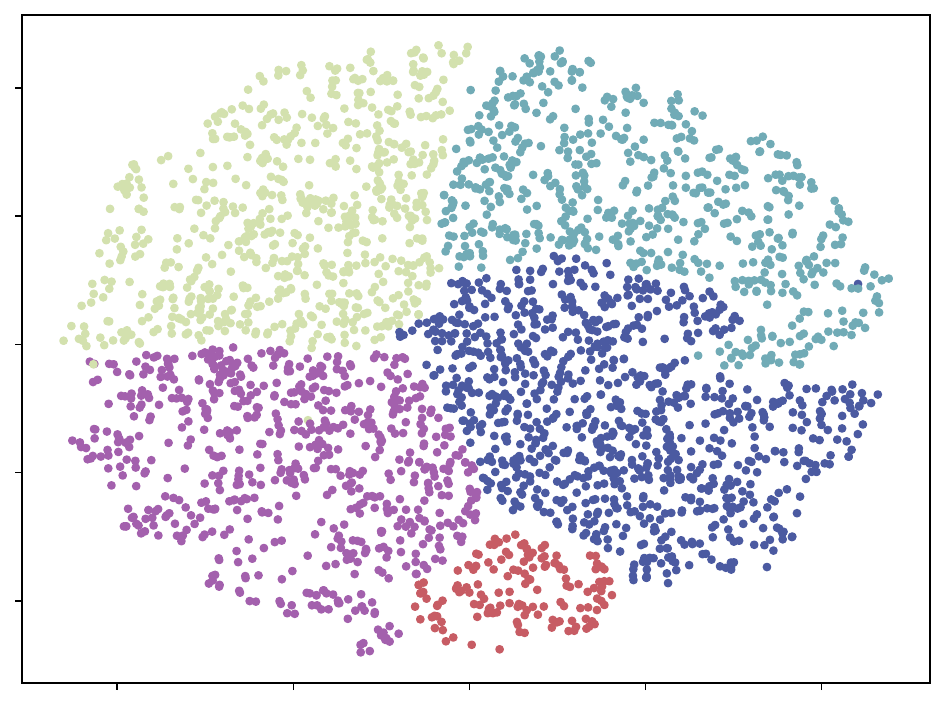}%
            \label{fig:rep_prev_conf}}
            \hfil
        \subfloat[fresh]{\includegraphics[width=0.11\textwidth]{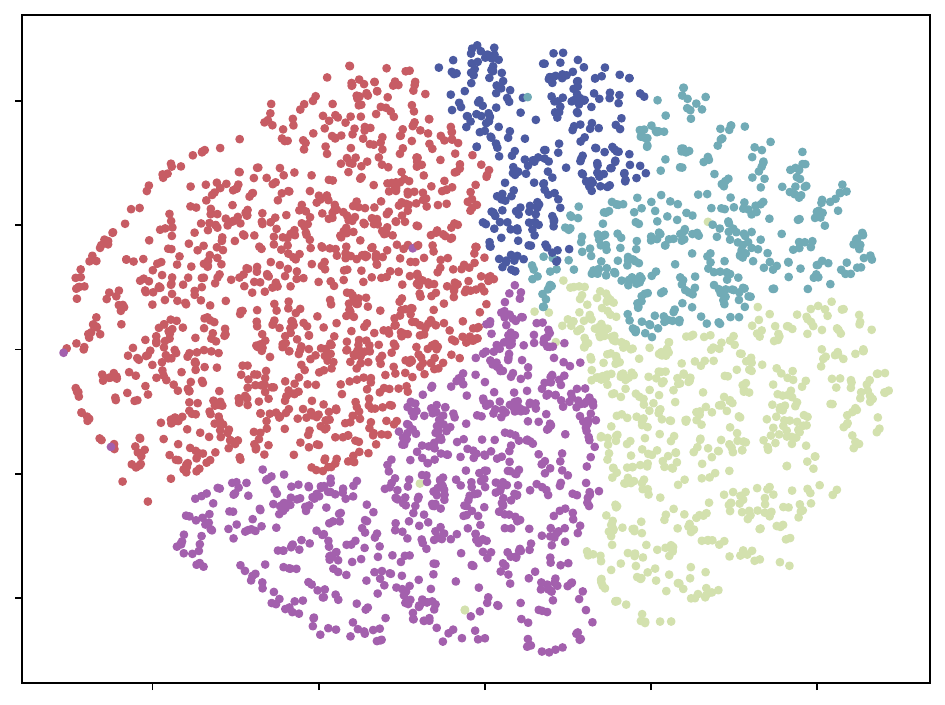}%
            \label{fig:rep_fresh_conf}}
              \hfil
        \subfloat[immed]{\includegraphics[width=0.11\textwidth]{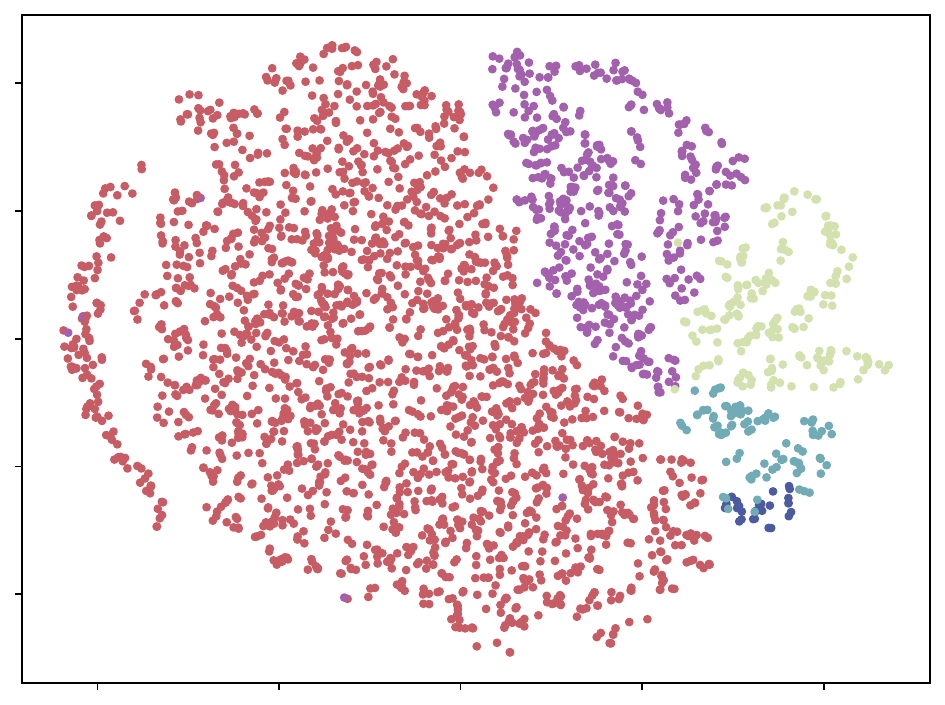}%
            \label{fig:rep_immed_conf}}
              \hfil
      \caption{Results of T-SNE visualization for representations from conformity perspective in CS. Various colors correspond to the binning labels assigned to the samples.}
    \label{fig:rep}
    \end{figure}

To validate that the disentangled perspectives learn distinct aspects of the features, we visualize the representations extracted from the conformity encoder using T-SNE \cite{scikit-learn}.
Figure \ref{fig:rep} illustrates a consistent and structured distribution of points across all figures. This indicates our model effectively captures group discrepancies across various binning labels for papers published at different times. Variations in the distribution of binning labels are evident across different categories due to differences in cumulative citations over time. Fresh papers published within the last five years, particularly immediate ones, tend to fall under the lowest accumulated citation label (red), dominating a significant area.
Despite this, discernible boundaries persist between distinct binning labels within all these categories. This suggests the model's capacity to generalize and transfer learned group differences to unseen samples, showcasing its potential applicability in real-world scenarios.

\section{Conclusion}
In this study, we propose a novel model named \textbf{DPPDCC} that disentangles the potential impacts of papers into diffusion, conformity, and contribution values. To better align with real-world conditions, we reformulate the problem by partitioning the datasets based on the observation time points without excluding lowly-cited papers. 
Further, DPPDCC comprises two principal components: Citation-aware GNN Encoder (CGE) and Popularity-aware Disentanglement Module (PDM). 
Given the dynamic heterogeneous graph of a target paper, CGE employs CompGAT to encode the comparative and co-cited/citing information between papers and utilizes Type-specific Attention Snapshot Readout to aggregate the information from each snapshot evolutionarily. 
Subsequently, PDM identifies amplifying effects in information diffusion and addresses the Matthew effect stemming from accumulated citations from a collective conformity perspective.
Finally, it constrains each perspective to address distinct aspects to preserve genuine contributions. Experimental results on three datasets demonstrate that DPPDCC significantly outperforms alternative approaches for previous, fresh, and immediate papers. 
Disentanglement visualization further demonstrates that DPPDCC can reliably perform predictions. 

\begin{acks}
    This work is supported by the National Natural Science Foundation of China (72204087, 72234005, 72104212), the Shanghai Planning Office of Philosophy and Social Science Youth Project (2022ET0001), "Chen Guang" project supported by Shanghai Municipal Education Commission and Shanghai Education Development Foundation (23CGA28), Shanghai Pujiang Program (23PJC030), the Fundamental Research Funds for the Central Universities, 2024 Innovation Evaluation Open Fund, Fudan University (CXPJ2024006). We also appreciate the constructive comments from the reviewers.
\end{acks}

\bibliographystyle{ACM-Reference-Format}
\balance
\bibliography{sample-base}

\appendix
\onecolumn
\section{Extended Related Work}
\subsection{Citation/Cascade prediction}
    Citation prediction is a vital sub-task of automatic academic evaluation as it enables the estimation of potential impact. It is closely examined within cascade prediction tasks, which share analogous graph structures and objectives \cite{zhou2021cascade}. The primary goal of cascade prediction is to anticipate a post's popularity based on user interactions. Similarly, in the context of citation networks, the paper can serve as the post, while the authors can serve as the users.
    Citation count prediction focuses on individual papers' content and distinctive features, heavily relying on comprehensive feature extraction techniques. In contrast, cascade prediction delves into the intricate relationships between various entities, including papers and authors, directly exploiting graph structure information.
    
    Their approaches can be broadly classified into three categories: stochastic models, feature-based models, and deep learning models.
    Stochastic models can predict future citation counts by fitting the curve of citation counts \cite{glanzel1995predictive}, following Zipf-Mandelbrot's law \cite{silagadze1997citations}. 
    Recently, machine learning models have shown promising results by utilizing manually extracted features from various meta-data (\textit{e.g.} content, author, venue) \cite{yan2011citation,yu2014citation,ruan2020predicting,dong2015cikm,guille2013cfeature,tatar2014cfeature2}. More presently, deep neural networks have dominated this task by applying advanced Natural Language Processing (NLP) and Computer Vision (CV) methods to extract abstractive representations from paper content \cite{abrishami2019predicting,huang2022fine,cohan2020specter,xue2023dgcbert}. For cascade prediction, graph embeddings, sequence models, and GNN models are employed to extract structural information from the underlying graph, and then temporal information is encoded with sequence models \cite{li2017deepcas,cao2017deephawkes,yang2022gtgcn,chen2022mucas,xu2022ccgl,cheng2023cassampling}. 
    
    However, most existing models for citation and cascade prediction exhibit sub-optimal performance. They fail to fully exploit the valuable information presented in paper content and scientific context within the citation network simultaneously. Furthermore, their strict data selection and random data splitting strategies may hinder their practicality. Many valuable samples will be filtered, and the temporal information within the dynamic context will be neglected. Thus, models in this setting are more prone to face distribution shift issues. 
    In contrast, our study revises the problem setting of impact prediction in the dynamic context \cite{he2023h2cgl, yan2024cody}.
    We ensure non-overlapping training and testing observation windows, retaining the real data distribution without sample filtering. We also utilize both content and dynamic context, introducing disentangled representation learning to provide a more practical impact estimation.

\subsection{Dynamic/Heterogeneous Graph Neural Networks}
    GNNs \cite{sperduti1997supervised,gori2005new,scarselli2008graph,wu2020comprehensive} are widely applied to handle non-Euclidean data like graphs. 
    Recently, research interest has surged in dynamic graphs, which incorporate temporal information, as well as in heterogeneous graphs, which involve multiple node types or edge types.
    These types of graphs are more prevalent in practical scenarios and encapsulate more intricate information compared to static homogeneous graphs, driving the formulation and refinement of diverse methodologies within the academic realm.
    
    The dynamic graph can be divided into multiple snapshots at different time points. Previous models like DGCN \cite{manessi2020dgcn}, Dysat \cite{sankar2020dysat}, and ROLAND \cite{you2022roland} typically encode the snapshots with static GCN as the structural encoder, followed by sequential models like Recurrent Neural Networks (RNN) or Transformer Encoder. 
    Moreover, certain models incorporate GCN into RNN models by configuring the parameters within RNN cells as GCN parameters \cite{pareja2020evolvegcn}.
    Within heterogeneous graphs, on one side, R-GCN serves as a prominent model with its separate message-passing architecture \cite{schlichtkrull2018rgcn}. It conducts independent message passing within different relations and aggregates them to update node representations. 
    On the other side, the relation-learning method transforms heterogeneous graphs into homogeneous ones, while retaining node and edge type indicators to preserve heterogeneous information \cite{hu2020hgt, you2022roland}. An example of this technique is HGT \cite{hu2020hgt}, which utilizes multi-head attention modules with specific parameters tailored to different node and edge types, effectively modeling the complexity of heterogeneous graphs.
    
    Citation networks can be regarded as dynamic and heterogeneous graphs as they evolve over time and encompass diverse entities such as papers, authors, and venues. Addressing the complexities within this network involves two primary challenges: (1) How to encode the temporal target-centric information within a single snapshot? (2) How to model distinct characteristics of information diffusion in citation networks? 
    To overcome these challenges, our study proposes a Citation-aware GNN Encoder that effectively models both dynamic and heterogeneous graphs of target papers. These graphs are constructed from the citation network annually and incorporate multiple metadata nodes beyond the papers themselves.
    (1) To encode the temporal target-centric information within a single snapshot, we employ a type-aware attention readout. It aggregates snapshot information considering the paper type and publication time.
    (2) To model distinct characteristics of information diffusion in citation networks, we propose a novel GCN module CompGAT. This module focuses on capturing comparisons and co-cited/citing information between papers, drawing inspiration from bibliometrics theories \cite{wu2019dindex}.

\subsection{Disentangled Representation Learning}
    Disentangled representation learning, aiming to discern and separate fundamental explanatory factors \cite{bengio2013replearning}, stands as a pivotal method within deep learning.
    It strives to produce resilient, manageable, and interpretable representations.
    Existing research predominantly focuses on the field of CV, often employing methods like Variational Autoencoders (VAEs) and Generative Adversarial Networks (GANs) to disentangle relevant latent factors \cite{ma2019disenrec,locatello2019challengingdisen,cai2019domaindisen,denton2017unsuperviseddisen,lee2021disenaug}.
    Researchers explore techniques to align distinct representations with specific factors through explicit and implicit guidance, employing approaches like direct labeling, loss penalties, and regularization.
    For instance, DR-GAN \cite{tran2017disentangled} utilizes an adversarial network to generate diverse facial poses, aiming to disentangle pose information from facial attributes. 
    In NLP, disentangled representation learning finds applications in various tasks like generation. For example, John et al. \cite{john2018disentangled} leverage VAE to disentangle text semantics from its writing style, enabling independent manipulation of text semantics and style. This separation allows precise control over content and style in text generation.
    
    Recently, there has been a growing interest in applying disentangled representation learning to graph learning tasks. Given that graph data encompasses both structural and attributed information, numerous new methodologies have emerged to address this complexity. In the feature space, approaches inspired by capsule networks have emerged to partition node features into multiple hidden channels for implicit disentangled modeling \cite{ma2019disengcn}. Some methods introduce additional constraints to enable effective separation \cite{liu2020ipgdn}. Moving into the structural space, subsequent models strive to factorize the input graphs into multiple subgraphs to facilitate distinct message-passing strategies \cite{yang2020factorgnn, li2021dgcl}. 
    Recent models have expanded their scope to isolate causal and biased information presented in both the graph structure and node features, with a particular emphasis on causal effects \cite{sui2022cal,fan2022disc}. These models commonly employ techniques to generate masks within the adjacency matrix and node features. This process aims to identify the crucial elements of the graph within the structural and attributed spaces.
    In addition, certain studies extend disentangled GNNs to specialized scenarios such as recommendation systems, handling intricate graph structures in heterogeneous or dynamic graphs \cite{wang2020disenhan, zhang2022dida, wen2022ddhgnn,zhang2023sild,zhang2023dyted}. Besides the masking strategy considering solely the graph structures and node features, they integrate classical methods like VAE and adversarial learning from CV and NLP to handle task-specific information from a broader perspective.
    
    In contrast to previous studies, our approach acknowledges that the potential impact within the citation network is influenced not only by the paper's contribution but also by factors related to its popularity. These popularity factors encompass the amplifying effect owing to the highly-cited paper nodes in information diffusion and the manifestation of the Matthew effect through collective conformity. 
    Following previous disentangled studies in recommendations \cite{zheng2021dice, chen2022poprec}, we consider the final predicted values as mixed results of multiple factors and introduce auxiliary tasks for separate encoding of different perspectives.
    Based on the dynamic heterogeneous graph, we disentangle the potential impact into diffusion, conformity, and contribution values, seeking a more reasonable estimation in explicit modeling. Specifically, we devise dedicated auxiliary tasks to extract values associated with these respective factors.

\section{Notation Table}
\begin{table*}[htbp]
  \centering
  \caption{The complete notation table of our paper.}
    \begin{tabular}{cl}
    \toprule
    Notation & \multicolumn{1}{c}{Description} \\
    \midrule
    $N$   & All predicted samples \\
    $G$   & Input dynamic heterogeneous graph \\
    $t$   & The $t$-th time point of the whole time window \\
    $G^t$ & The $t$-th subgraph of the input graph      \\
    $\epsilon^t$, $\nu^t$ &  Set of nodes and edges at $t$-th time point  \\
    $\phi$, $\varphi$ & Type map function of nodes and edges \\
    $U$, $R$ & Set of node types and edge types \\
    $p$   & The paper node type \\
    \midrule
    $\bm{h}_{ti, u}^{(l)}$ & The $i$-th node hidden representation of type $u$ at $t$-th time step in $l$-th layer \\
    $\bm{H}_s$ & Snapshot representation matrix containing all $h_{ti, s}$ of the $i$-th sample   \\
    $AGG$ & Aggregation functions of update representations from multiple relationships (edges) \\
    $f_r$ &  Message passing method (module) of the edge $r$   \\
    $g_r$ & Subgraph of the edge $r$     \\
    $\bm{h}_{r_{src}}$, $\bm{h}_{r_{dst}}$ & The source and target node representations of the edge r \\
    $\bm{W}$ & The model parameters \\
    $\bm{C}$ & Normalized co-cited/citing strength matrix \\
    $\bm{o}_i$ & The final encoded representation of the $i$-th target sample before disentanglement  \\
    $\bm{o}_i^{dif}$, $\bm{o}_i^{con}$ &  The representation of the $i$-th sample's diffusion and conformity view   \\
    $\bm{z}_i^{ori \vee pos \vee neg}$ & The $i$-th sample's projected representation of the original, positive and negative view \\
    \midrule
    $\mathcal{L}_{dif}$ & Objective to disentangle value of diffusion view \\
    $\mathcal{L}_{con}$ & Objective to disentangle value of conformity view \\
    $\mathcal{L}_{ort}$ & Orthogonal constraint to ensure the non-overlapping information of disentangled views \\
    $\mathcal{L}_{reg}$ & Objective to predict the citation increments as the main task \\
    \midrule
    $L$   & The number of stacked graph encoder \\
    $\lambda$ & The importance of co-cited/citing strengths \\
    $\beta$ & The weight of the whole disentanglement task \\
    \bottomrule
    \end{tabular}%
  \label{tab:notation}%
\end{table*}%

\section{Complexity Analysis}
The primary complexity lies in the graph encoder component of our model, CGE. Within each layer of CGE, multiple R-GCNs operate on snapshots captured at different time points, alongside a snapshot representation updater and a multi-layer transformer encoder. For edges apart from "cites" and "is cited by" in R-GCNs, a simple MLP ($W^u \in \mathbb{R}^{d \times d}$) serves as learnable parameters for GIN, with a complexity of $\mathcal{O}\left(LT(|R|-2)(|\nu^t_{r_{src}}|d^2 + |\epsilon^t_r|d)\right)$. For "cites" and "is cited by" edges, the backbone of CompGAT, GATv2, incurs a complexity of $\mathcal{O}\left(LT(|\nu^t_p|d^2 + |\epsilon^t_p|d)\right)$. Due to the modeling of concatenated source and target representations, it further requires $\mathcal{O}\left(LT|\epsilon^t_p|d^2)\right)$ time, resulting in an overall complexity of $\mathcal{O}\left(LT(|\nu^t_p| + |\epsilon^t_p|)d^2\right)$. The snapshot representation updater, functioning as a cross-attention module, operates with a complexity of $\mathcal{O}\left(LT(|\nu^t_p|d + d^2)\right)$, while the temporal encoder exhibits a complexity of $\mathcal{O}\left(L(T^2d + Td^2)\right)$. These modules can be readily disregarded compared to R-GCNs. For PDE, which solely models the final target representation using multiple MLPs ($\mathcal{O}\left(d^2\right)$), its contribution can also be omitted. Thus, the resulting complexity of DPPDCC is $\mathcal{O}\left(LT(|\nu^t|d^2 + |\epsilon^t_p|d^2 + |\epsilon^t|d)\right)$. The final time complexity depends on the sampling strategy for graph size and the sparsity of \textit{paper} node subgraphs. Since citation graphs are sparse and only a small portion of the entire graphs is sampled, computations are confined mainly within the limited \textit{paper} subgraphs. 
Therefore, DPPDCC is comparable to other dynamic heterogeneous GNNs.

\section{Experimental Details}
\subsection{Datasets}
\begin{table*}[htbp]
  \centering
  \setstretch{0.8}
  \caption{Detailed statistics for ground-truth samples, nodes, and edges of the citation network, as well as sample information of training, validation, and testing sets.}
    \begin{tabular}{c|l|r|r|r}
    \toprule
    \multicolumn{2}{c|}{dataset} & \multicolumn{1}{c|}{CS} & \multicolumn{1}{c|}{CHM} & \multicolumn{1}{c}{PSY} \\
    \midrule
    \multicolumn{2}{c|}{ground-truth sample} & 2,513,197  & 1,818,138  & 1,857,277  \\
    \midrule
    \multirow{4}[2]{*}{node} & paper & 1,628,853  & 1,376,599  & 1,297,771  \\
          & author & 1,598,925  & 1,946,073  & 1,585,595  \\
          & venue & 12,524  & 8,389  & 13,775  \\
          & time  & 147   & 187   & 202  \\
    \midrule
    \multirow{4}[2]{*}{edge} & cite  & 11,534,431  & 10,382,698  & 13,401,112  \\
          & write & 5,123,460  & 6,355,630  & 4,813,135  \\
          & publish & 1,566,442  & 1,318,158  & 1,230,440  \\
          & have  & 1,628,853  & 1,376,599  & 1,297,771  \\
    \midrule
    \multirow{4}[4]{*}{sample} & train (2012) & 183,105  & 227,304  & 219,499  \\
          & val (2014) & 224,086  & 256,403  & 250,229  \\
          & test (2017) & 300,000  & 300,000  & 300,000  \\
\cmidrule{2-5}          & qualified pool & 1,112,611  & 1,093,660  & 1,015,579  \\
    \bottomrule
    \end{tabular}%
  \label{tab:data_full}%
\end{table*}%

Our dataset is constructed using S2AG \cite{kinney2023s2orcnew}, a comprehensive repository housing roughly 100 million scientific publications, offering extensive metadata and unique identifiers. To enhance experimental efficiency and accommodate the diverse characteristics across domains, we organize these publications into subsets based on their respective fields of study.
To establish the global citation network, we locate English papers with complete metadata (title, abstract, author IDs, and venue ID) or those of high impact despite lacking venue information. From the dataset, considering the data quality and the disciplinary differences, we select three fields: computer science (CS), chemistry (CHM), and psychology (PSY). 
Their total number of papers and the completeness of their metadata are the highest among all fields. 
Additionally, the differences in citation patterns resulting from varying research paradigms, such as variations in citation half-lives \cite{mendoza2021citepat}, render these fields appropriate representative samples for demonstrating citation differences across disciplines.
Relevant statistics are summarized in Table \ref{tab:data_full}.

To select samples for testing, we target papers published before the test observation point. These papers should be written in English, have complete metadata, and include at least one reference without imposing constraints on the number of citations received to maintain the integrity of the data distribution.
From the eligible pool, we randomly sampled 300,000 papers to compose the test set. 
Additionally, we gather all the associated metadata for these papers, encompassing titles, abstracts, authors, venues, and publication times. 
To imitate the practical scenario, the training observation point is set 5 years before the test observation point (equal to the observation time window $T$ and predicted interval $\Delta$) for non-overlapping. The validation observation point precedes the test observation point by 3 years. 
It is crucial to underscore that though predicted samples in the training set reappear in the test set, their input graphs and predicted values undergo significant variations due to dynamic contextual changes. This intrinsic challenge renders the task both complex and critical. Moreover, we meticulously categorize the papers in the results for thorough and detailed analysis.

\begin{table*}[htbp]
  \centering
  \setstretch{0.8}
  \caption{Detailed statistics of single input heterogeneous subgraph for DPPDCC.}
    \begin{tabular}{c|l|rrr|rrr|rrr}
    \toprule
    \multicolumn{2}{c|}{dataset} & \multicolumn{3}{c|}{CS} & \multicolumn{3}{c|}{CHM} & \multicolumn{3}{c}{PSY} \\
    \midrule
    \multicolumn{2}{c|}{phase} & \multicolumn{1}{c}{train} & \multicolumn{1}{c}{val} & \multicolumn{1}{c|}{test} & \multicolumn{1}{c}{train} & \multicolumn{1}{c}{val} & \multicolumn{1}{c|}{test} & \multicolumn{1}{c}{train} & \multicolumn{1}{c}{val} & \multicolumn{1}{c}{test} \\
    \midrule
    \multirow{4}[2]{*}{nodes} & paper & 204   & 222   & 257   & 149   & 187   & 254   & 361   & 401   & 460  \\
          & author & 446   & 501   & 613   & 505   & 646   & 907   & 1015  & 1173  & 1431  \\
          & venue & 72    & 80    & 92    & 35    & 41    & 50    & 92    & 104   & 123  \\
          & time  & 23    & 24    & 26    & 22    & 24    & 27    & 32    & 34    & 37  \\
    \midrule
    \multirow{3}[2]{*}{edges} & cites & 740   & 827   & 1016  & 480   & 662   & 1036  & 1607  & 1808  & 2085  \\
          & writes & 599   & 674   & 827   & 673   & 864   & 1218  & 1388  & 1598  & 1933  \\
          & publishes & 165   & 183   & 216   & 135   & 168   & 228   & 311   & 350   & 408  \\
    \bottomrule
    \end{tabular}%
  \label{tab:input}%
\end{table*}%

In Table \ref{tab:input}, we present detailed statistics of the input data for DPPDCC to demonstrate the sufficiency of information within multi-hop heterogeneous subgraphs. For simplicity, we average the node and edge counts by time and sample. It is evident that the Psychology dataset is the densest, while the Computer Science and Chemistry datasets are comparable.

\subsection{Descriptive Statistics}
To validate the rationale behind the disentangled factors used in our work, we visualize the distribution of citation increments between papers potentially influenced by corresponding popularity factors and other normal papers. For the information diffusion effect, it is posited that documents cited by highly-cited papers are more likely to attract attention from other papers during dissemination, thereby accruing more citations. Conversely, for the collective conformity effect, it is argued that the cumulative citations received by a paper can reflect its partial reputation and consequently increase its likelihood of being cited. Highly-cited papers mentioned in both contexts are classified using a Pareto distribution method: papers providing over 80\% of the total cumulative citations are deemed highly-cited. The papers cited by these highly-cited papers represent focused samples of the information diffusion effect, while the highly-cited papers themselves serve as focused samples of the collective conformity effect.

\begin{figure*}[]
      \centering
        \centering
        \subfloat[CS]{\includegraphics[width=0.25\textwidth]{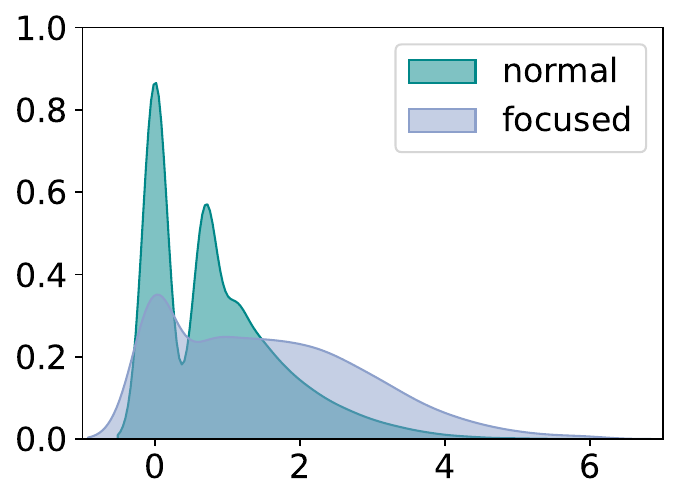}%
            \label{fig:stat_cited_cs}}
            \hfil
        \subfloat[CHM]{\includegraphics[width=0.25\textwidth]{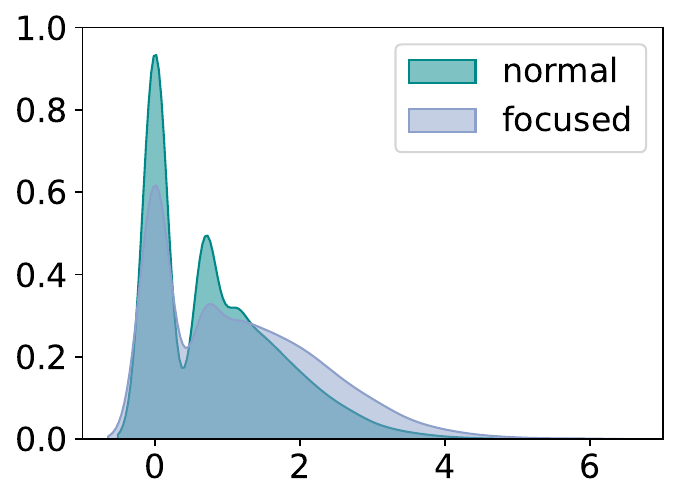}%
            \label{fig:stat_cited_chm}}
            \hfil
        \subfloat[PSY]{\includegraphics[width=0.25\textwidth]{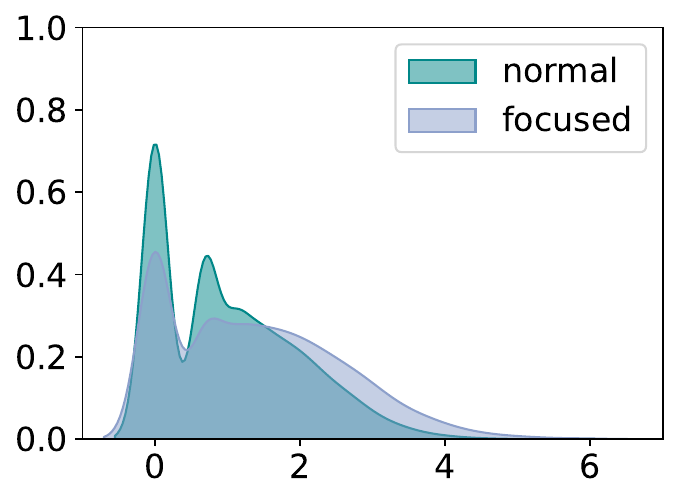}%
            \label{fig:stat_cited_psy}}
              \hfil
      \caption{Citation increment distribution of normal samples and focused samples cited by highly-cited papers for diffusion perspective.}
    \label{fig:stat_cited}
    \end{figure*}

(1) In Figure \ref{fig:stat_cited}, the significant deviation among different categories is evident, with papers cited by highly-cited ones generally receiving more citations. Notably, the CS dataset exhibits the most distinct pattern, with the left peak nearly disappearing, resulting in an overall distribution closer to a normal distribution. The model aims to extract these diffusion-related factors to model the information diffusion effect independently.

\begin{figure*}[]
      \centering
        \centering
        \subfloat[CS]{\includegraphics[width=0.25\textwidth]{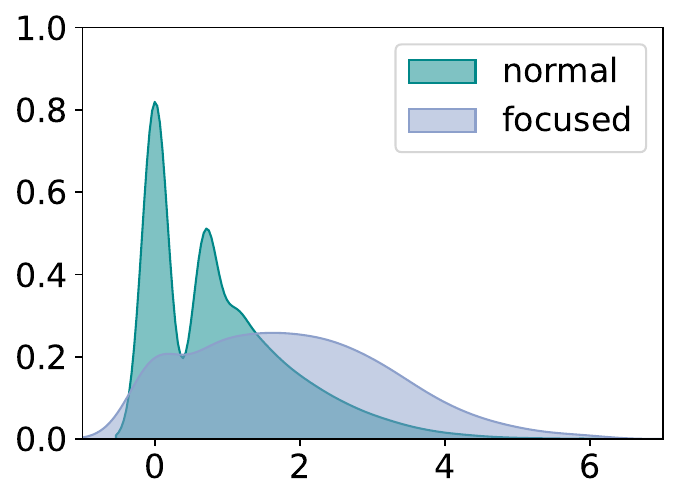}%
            \label{fig:stat_accum_cs}}
            \hfil
        \subfloat[CHM]{\includegraphics[width=0.25\textwidth]{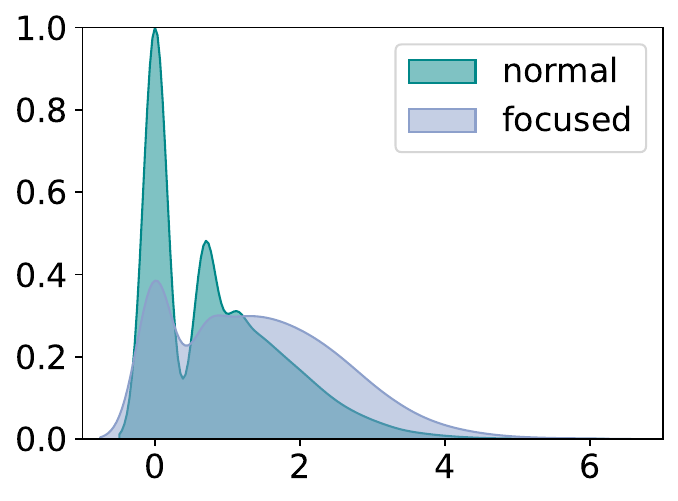}%
            \label{fig:stat_accum_chm}}
            \hfil
        \subfloat[PSY]{\includegraphics[width=0.25\textwidth]{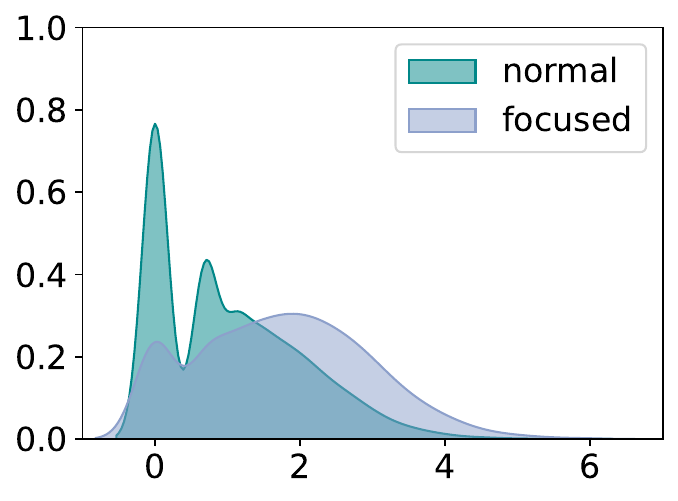}%
            \label{fig:stat_accum_psy}}
              \hfil
      \caption{Citation increment distribution of normal samples and focused highly-cited samples for conformity perspective.}
    \label{fig:stat_accum}
    \end{figure*}

(2) Figure \ref{fig:stat_accum} illustrates the difference in attracting new citations between papers with accumulated citations and those with fewer citations, indicating that the collective conformity effect is more pronounced than the information diffusion effect. Notably, the distribution of focused samples in the CS dataset has essentially normalized, while other datasets also display significant peak differences. The model is expected to separately model the conformity effect, enabling it to focus more on aspects unrelated to existing cumulative citations and align more closely with contribution-related information.

\subsection{Baseline Models}
The baseline models are mainly divided into citation/cascade prediction models, dynamic GNN models, and disentangled GNN models:

\textbf{Citation/Cascade prediction models:}
We apply classic and recent citation/cascade prediction models to estimate the potential impacts of papers based on content, citation graphs, or citation cascade graphs. 
\begin{itemize}
    \item[-] \textbf{SciBERT} \cite{beltagy2019scibert} is a BERT-based model pretrained on scientific corpus. We fine-tune it to encode paper titles and abstracts, predicting with the [CLS] token for citation prediction.
    \item[-] \textbf{SPECTER2} \cite{cohan2020specter} generates document-level embeddings of scientific documents by pretraining a BERT-based model on relatedness signals extracted from the citation graph. Instead of fine-tuning the model's adapters, we fully fine-tune the basic SPECTER2 for better performance.
    \item[-] \textbf{HINTS} \cite{jiang2021hints} tackles the cold-start problem by employing R-GCN on pseudo meta-data heterogeneous graphs. It then feeds the node (paper) representations into GRU for sequential modeling and employs a stochastic model for predictions. 
    To integrate content information from PLMs, we replace node ID embeddings and adjust the model to encode subgraphs instead of the entire graph, mitigating memory issues.
    \item[-] \textbf{CasSampling} \cite{cheng2023cassampling} employs an importance-aware sampling method to extract critical components of cascade graphs based on outdegree, retaining the time series of the global graphs as node features. Subsequently, an attention aggregator is used to encode the global-level time flow.
    \item[-] \textbf{H$^2$CGL} \cite{cheng2023cassampling}, the state-of-the-art citation prediction model, shares a similar dynamic setting and heterogeneous GNN framework with our approach. It encodes the dynamics of the citation network hierarchically and designs message-passing modules that consider the unique characteristics of citation networks. Additionally, it incorporates graph contrastive learning to refine representations, encouraging the model to encode potential citation information unrelated to specific topics.
\end{itemize}

\textbf{Dynamic GNN models:}
We employ recent dynamic GNN models to extract paper representations from the dynamic graphs collected in the same way as our model, taking the potential impact prediction as a graph regression task. Similar to HINTS, we adapt these models to concentrate on target-centric subgraphs rather than the entire graph.
\begin{itemize}
    \item[-] \textbf{EvolveGCN} \cite{pareja2020evolvegcn} integrates RNN structure within GNN modules. It regards graph structures as hidden states within the GRU. It adopts GRU parameters as GCN parameters. 
    \item[-] \textbf{Dysat} \cite{sankar2020dysat} incorporates multiple self-attention modules to encode structural and temporal information. It first applies a shared GAT to encode subgraphs separately across different years. Subsequently, a vanilla transformer encoder models the temporal information of the graph representation sequence.
    \item[-] \textbf{ROLAND} \cite{you2022roland} is a novel dynamic heterogeneous GCN model that inherits static GCN methods. It retains the hierarchical information of GCNs' different layers and then applies a GRU-like updater to iteratively update representations.
\end{itemize}

\textbf{Disentangled GNN models:}
We apply recent disentangled GNN models to extract critical representations from the multi-hop subgraphs. 
\begin{itemize}
    \item[-] \textbf{DisenGCN} \cite{ma2019disengcn} partitions the feature vector into K channels and uses neighborhood routing from capsule networks for disentanglement.
    \item[-] \textbf{DisenHAN} \cite{wang2020disenhan} adapts DisenGCN to heterogeneous graphs for recommendation. It employs DisenGCNs in different homogeneous graphs for intra-aggregation. It then employs the attention mechanism to aggregate information from different relationships under various channels.
    \item[-] \textbf{CAL} \cite{sui2022cal} employs attention modules to estimate the causal and trivial masks for structures and attributes. It encodes them with specific GCNs separately and applies causal intervention by randomly swapping the trivial part of the whole embeddings.
    \item[-] \textbf{DisC} \cite{fan2022disc} first estimates the causal/bias masks and then employs causal/bias-aware loss functions. It further generates counterfactual unbiased samples in the embedding space by swapping the biased part of the representations.
    \item[-] \textbf{DIDA} \cite{zhang2022dida} discovers invariant patterns within dynamic graphs by generating soft masks using the self-attention mechanism. It then applies random variant pattern swapping in both temporal and spatial dimensions as the causal intervention.
    \item[-] \textbf{DyTed} \cite{zhang2023dyted} designs a temporal-clips contrastive learning task along with a structural contrastive learning task to effectively disentangle time-invariant and time-varying representations. It further employs an adversarial learning framework with a disentanglement-aware discriminator for enhancement. EGCN and Dysat are utilized as the backbone dynamic GNNs of DyTed, with the best performance reported using Dysat.
    \item[-] \textbf{SILD} \cite{zhang2023sild} addresses distribution shifts in both dynamic and spectral domains. It builds upon DIDA with spectral transformation and proposes a disentangled spectrum mask to capture and utilize invariant and variant spectral patterns.
\end{itemize}

\subsection{Implementation Details}
All baselines and our proposed model are implemented using PyTorch and DGL \cite{wang2019dgl}/PyG \cite{Fey/Lenssen/2019pyg}. They are trained on an NVIDIA A800 80GB GPU and optimized with the Adam optimizer \cite{kingma2014adam}. For the baselines, we directly use or refer to the official source codes. Experiments are conducted with three random seeds, and the averaged results are reported. Models are developed using the training and validation sets, with the final model selected based on the best main metric on the validation set. Each dataset is divided into previous papers (included in the training set), fresh papers (new additions in the test set), and immediate papers (published at the test time point) for detailed analysis. 
We follow the recommended settings from the original papers or open-source codes for all baselines.
In our model across all datasets, the hop order $k$ is set to 2, corresponding top limits $K_1$ and $K_2$ are set to 100 and 20, the learning rate is set to 1e-4, the number of layers $L$ is set to 4, the co-cited/citing weight $\lambda$ and the disentanglement weight $\beta$ are set to 0.5, and the categories of equal frequency bins $M$ are set to 5. Additionally, the batch size is set to 64 for CS and CHM, and 32 for PSY, and the hidden dimension $d$ is set to 128 for CS and PSY, and 192 for CHM.

\subsection{Evaluation Metric}
In our evaluation, we utilize two metrics: MALE (Mean Absolute Logarithmic Error) and LogR$^2$ (Logarithmic R-squared). MALE calculates the absolute error between the target and predicted values post-logarithm transformation, reflecting the direct predictive capability:
\begin{equation}
    MALE = \frac {1}{N} \sum_{i=1}^{N}{|y_i - \hat{y}_i|}
\end{equation}
Regarding LogR$^2$, it evaluates the coefficient of determination, calculated as the proportion of the dependent variable's variation predictable from the independent variable, following logarithm transformation for both variables:
\begin{equation}
    LogR^2 = 1 - \frac{\sum_{i=1}^{N}{( y_i - \hat{y}_i)^2}}{\sum_{i=1}^{N}{(y_i - \frac {1}{N} \sum_{i=1}^{N} y_i)^2}}
\end{equation}
LogR$^2$ assumes a negative value when the model's predictions fall short of the data's mean value, indicating an inability to capture the underlying trend within the dataset. In our problem context, mirroring practical scenarios within dynamic contexts that require extrapolative abilities, encountering negative values is both frequent and expected due to inherent complexities. Consequently, by integrating MALE and LogR$^2$, we holistically gauge the model's accuracy in direct predictions for individual samples and its aptness in capturing broader trends across groups.

\section{Complete Experimental Results}
\label{subsec:disen_visual}
\begin{table*}[htbp]
  \centering
  \caption{The complete experimental results of total samples with standard deviations.}
    \begin{tabular}{l|rr|rr|rr}
    \toprule
    \multicolumn{1}{c|}{\multirow{2}[4]{*}{model}} & \multicolumn{2}{c|}{CS} & \multicolumn{2}{c|}{CHM} & \multicolumn{2}{c}{PSY} \\
\cmidrule{2-7}          & \multicolumn{1}{c}{MALE↓} & \multicolumn{1}{c|}{LogR2↑} & \multicolumn{1}{c}{MALE↓} & \multicolumn{1}{c|}{LogR2↑} & \multicolumn{1}{c}{MALE↓} & \multicolumn{1}{c}{LogR2↑} \\
    \midrule
    SciBERT & 0.6904\textpm.007 & 0.3370\textpm.016 & 0.5868\textpm.030 & 0.3887\textpm.054 & 0.6131\textpm.019 & 0.4761\textpm.034 \\
    SPECTER2 & 0.7482\textpm.027 & 0.2337\textpm.058 & 0.5832\textpm.013 & 0.3906\textpm.032 & 0.5935\textpm.020 & 0.5082\textpm.031 \\
    HINTS & 0.9022\textpm.108 & -0.1227\textpm.231 & 0.8265\textpm.074 & -0.1283\textpm.160 & 0.8529\textpm.081 & -0.0198\textpm.204 \\
    CasSampling & 0.5061\textpm.003 & 0.6112\textpm.007 & 0.4679\textpm.002 & 0.6127\textpm.003 & 0.5196\textpm.002 & 0.6200\textpm.006 \\
    H2CGL & 0.4579\textpm.007 & 0.6812\textpm.014 & 0.4337\textpm.003 & 0.6573\textpm.010 & 0.4727\textpm.005 & 0.6836\textpm.006 \\
    \midrule
    EGCN  & 0.8051\textpm.009 & 0.1296\textpm.009 & 0.6381\textpm.007 & 0.2826\textpm.020 & 0.7339\textpm.012 & 0.2530\textpm.021 \\
    Dysat & 0.6297\textpm.006 & 0.4242\textpm.015 & 0.5412\textpm.009 & 0.4705\textpm.017 & 0.6207\textpm.047 & 0.4354\textpm.096 \\
    ROLAND & 0.5978\textpm.033 & 0.4911\textpm.048 & 0.5437\textpm.037 & 0.4495\textpm.073 & 0.5521\textpm.014 & 0.5718\textpm.017 \\
    \midrule
    DisenGCN & 0.8321\textpm.010 & 0.1026\textpm.030 & 0.6343\textpm.006 & 0.2897\textpm.016 & 0.6689\textpm.011 & 0.3746\textpm.018 \\
    DisenHAN & 0.8221\textpm.006 & -0.0488\textpm.038 & 0.8652\textpm.181 & -0.1110\textpm.355 & 0.8766\textpm.191 & -0.1186\textpm.552 \\
    CAL   & 0.7094\textpm.018 & 0.1859\textpm.222 & 0.5733\textpm.015 & 0.4166\textpm.026 & 0.5535\textpm.009 & 0.5783\textpm.013 \\
    DisC  & 0.6733\textpm.006 & 0.3496\textpm.013 & 0.5609\textpm.031 & 0.4259\textpm.065 & 0.5592\textpm.021 & 0.5606\textpm.035 \\
    DIDA  & 0.6203\textpm.034 & 0.4449\textpm.045 & 0.5108\textpm.016 & 0.5354\textpm.035 & 0.5723\textpm.022 & 0.5328\textpm.034 \\
    \midrule
    DyTed (EGCN) & 0.8101\textpm.008 & 0.1235\textpm.008 & 0.6411\textpm.007 & 0.2465\textpm.006 & 0.7896\textpm.009 & 0.1478\textpm.025 \\
    DyTed (Dysat) & 0.6820\textpm.035 & 0.2426\textpm.138 & 0.5386\textpm.010 & 0.4602\textpm.013 & 0.5838\textpm.001 & 0.5029\textpm.001 \\
    SILD  & 0.5836\textpm.011 & 0.4977\textpm.031 & 0.5009\textpm.003 & 0.5629\textpm.004 & 0.5448\textpm.009 & 0.5889\textpm.016 \\
    \midrule
    DPPDCC & 0.4432\textpm.005 & 0.6938\textpm.008 & 0.4264\textpm.002 & 0.6667\textpm.005 & 0.4587\textpm.002 & 0.6969\textpm.002 \\
    \#improve (\%) & 3.21$^*$  & 1.85$^*$  & 1.69$^*$  & 1.43$^*$  & 2.96$^*$  & 1.94$^*$  \\
    \bottomrule
    \end{tabular}%
  \label{tab:complete}%
\end{table*}%

We run the experiments with three random seeds and report both the averaged results and standard deviations of all models. As shown in Figure \ref{tab:complete}, our proposed model DPPDCC consistently outperforms others with minimal variation, highlighting its capacity and stability.

\section{Complexity Experiment}

\begin{table}[htbp]
  \centering
  \caption{The input data, parameters, and inference time of selected baselines and DPPDCC.}
    \begin{tabular}{c|l|cc|rr|rr|rr}
    \toprule
    \multirow{2}[4]{*}{Type} & \multicolumn{1}{c|}{\multirow{2}[4]{*}{model}} & \multicolumn{2}{c|}{Input Data} & \multicolumn{2}{c|}{CS} & \multicolumn{2}{c|}{CHM} & \multicolumn{2}{c}{PSY} \\
\cmidrule{3-10}          &       & Dynamic & Hetero & \multicolumn{1}{c}{params} & \multicolumn{1}{c|}{inference} & \multicolumn{1}{c}{params} & \multicolumn{1}{c|}{inference} & \multicolumn{1}{c}{params} & \multicolumn{1}{c}{inference} \\
    \midrule
    PLM   & SciBERT & $\times$     & $\times$     & 111.1M & 2.91s & 111.1M & 2.92s & 111.1M & 2.92s \\
    \midrule
    Cascade
          & CasSampling & $\times$     & $\times$     & 0.2M  & 0.02s & 0.2M  & 0.02s & 0.2M  & 0.02s \\
    \midrule
    \multirow{2}[2]{*}{Dynamic} & Dysat & \checkmark     & $\times$     & 0.2M  & 0.66s & 0.2M  & 0.76s & 0.2M  & 1.24s \\
          & ROLAND & \checkmark     & \checkmark     & 0.5M  & 4.76s & 0.5M  & 6.59s & 0.5M  & 8.71s \\
    \midrule
    \multirow{4}[2]{*}{Disentangled} & CAL   & $\times$     & $\times$     & 0.3M  & 0.07s & 0.3M  & 0.07s & 0.3M  & 0.11s \\
          & DIDA  & \checkmark     & $\times$     & 1.4M  & 2.26s & 1.4M  & 2.55s & 1.4M  & 4.89s \\
          & DyTed & \checkmark     & $\times$     & 2.1M  & 0.79s & 2.1M  & 0.78s & 2.1M  & 1.37s \\
          & SILD  & \checkmark     & $\times$     & 2.2M  & 1.31s & 2.2M  & 1.42s & 2.2M  & 2.40s \\
    \midrule
    ours  & DPPDCC & \checkmark     & \checkmark     & 14.3M & 4.29s & 31.0M & 6.06s & 14.3M & 7.67s \\
    \bottomrule
    \end{tabular}%
  \label{tab:complexity}%
\end{table}%

Our model's complexity is primarily influenced by the R-GCN architecture, which requires distinct message-passing. During training, proposed contrastive learning involves additional encoding of multi-view graphs, but this step can be omitted during inference. Notably, our model converges quickly, requiring only 5 epochs compared to 20 epochs for other baselines, ensuring comparable training time. Given that we train and test on datasets with approximately 0.2 million samples, it ensures comparable coverage. Figure \ref{tab:complexity} demonstrates that our model has comparable space occupancy and time complexity to the baseline models.

\section{Extended Visualizations}
\subsection{Weighting Score Visualization}
    \begin{figure*}[]
      \centering
        \centering
        \subfloat[][$e^{(1)} = f(c)$]{
        \begin{tabular}{c} 
            \includegraphics[width=0.21\textwidth]{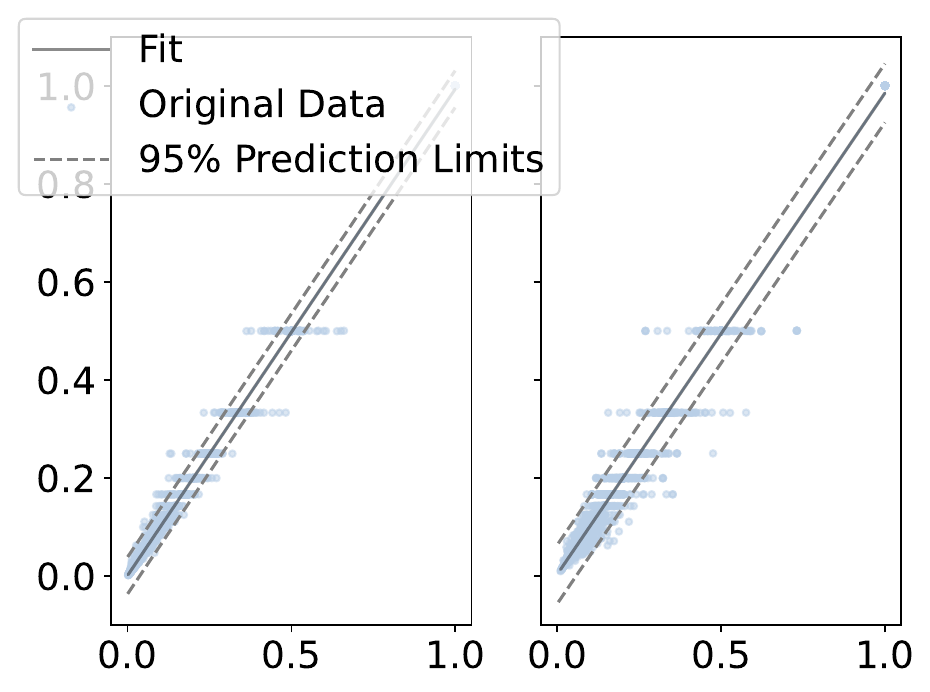} \\
            (8.56/21.00)
        \end{tabular}{c}
        \label{fig:attn_ac0}
            }
            \hfil
        \subfloat[][$e^{(2)} = f(c)$]{
        \begin{tabular}{c} 
            \includegraphics[width=0.21\textwidth]{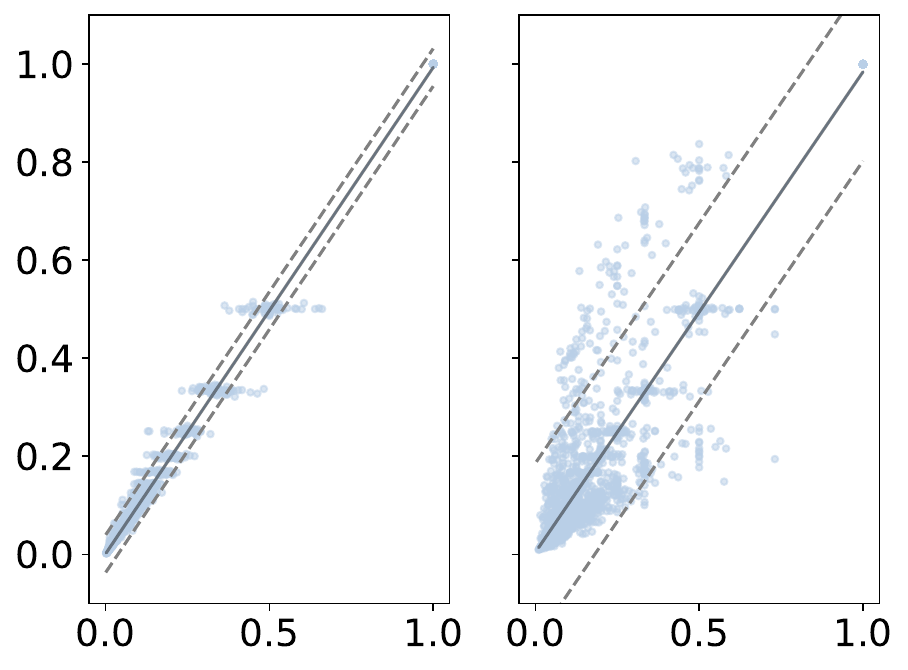} \\
            (8.97/184.96)
        \end{tabular}
        \label{fig:attn_ac1}
            }
            \hfil
        \subfloat[][$e^{(3)} = f(c)$]{
        \begin{tabular}{c} 
            \includegraphics[width=0.21\textwidth]{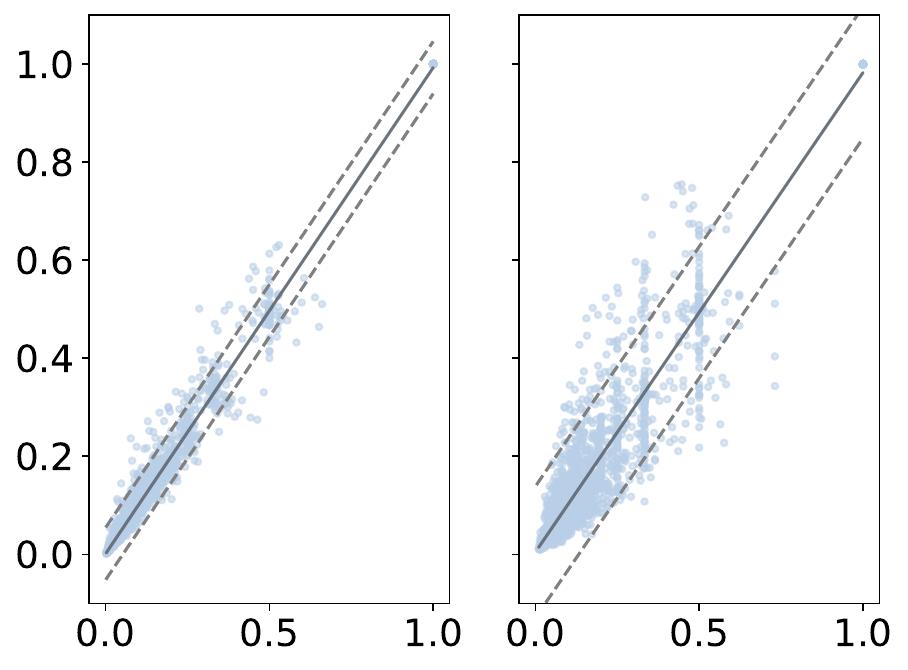} \\
            (19.23/106.80)
        \end{tabular}
        \label{fig:attn_ac2}
            }
              \hfil
        \subfloat[][$e^{(4)} = f(c)$ \\ ]{
        \begin{tabular}{c} 
            \includegraphics[width=0.21\textwidth]{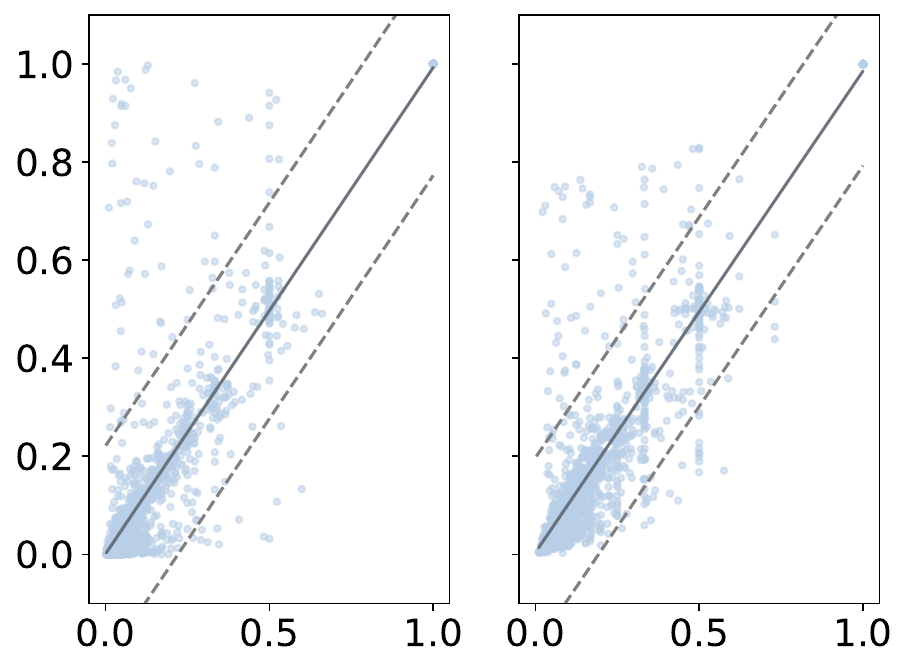} \\
            (373.88/240.24)
        \end{tabular}
        \label{fig:attn_ac3}
            }
              \hfil
              
        \subfloat[$\alpha^{(1)} = f(e^{(1)})$]{
        \begin{tabular}{c} 
            \includegraphics[width=0.21\textwidth]{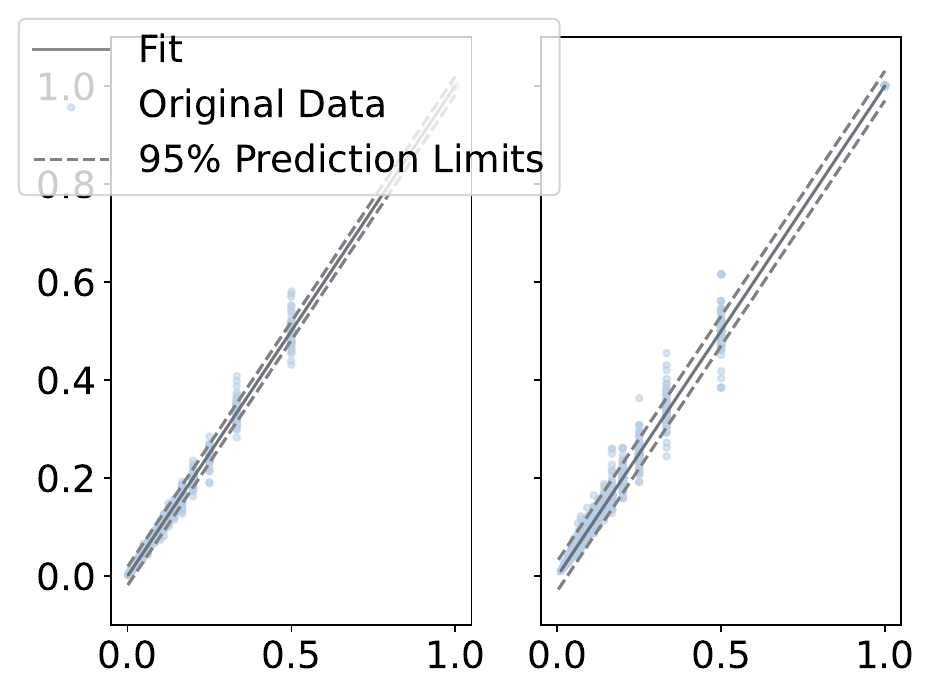} \\ 
            (1.98/5.42)
        \end{tabular}
        \label{fig:attn_sa0}
            }
            \hfil
        \subfloat[$\alpha^{(2)} = f(e^{(2)})$]{
        \begin{tabular}{c} 
            \includegraphics[width=0.21\textwidth]{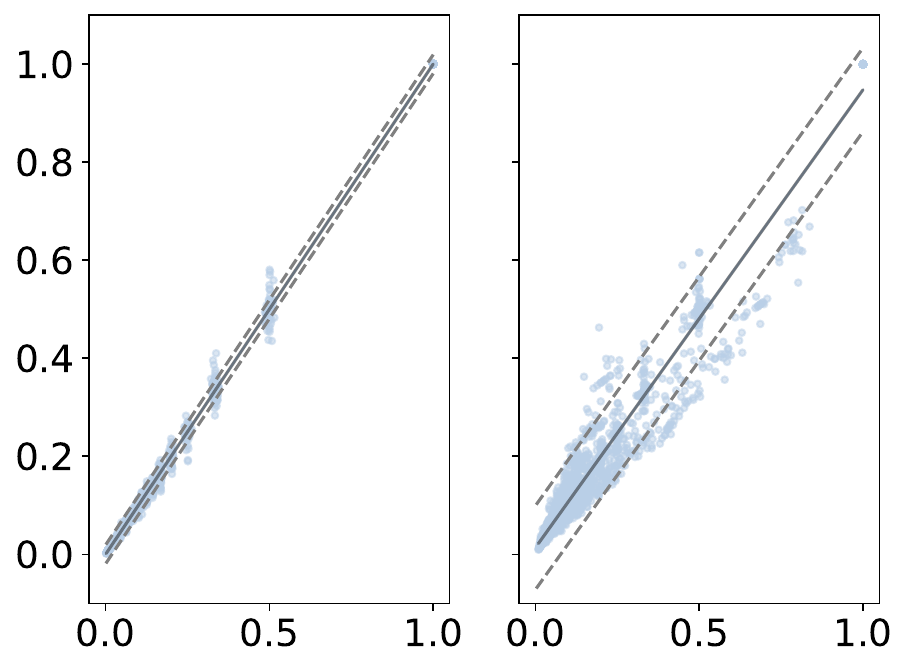} \\ 
            (2.06/43.91)
        \end{tabular}
        \label{fig:attn_sa1}
            }
            \hfil
        \subfloat[$\alpha^{(3)} = f(e^{(3)})$]{
        \begin{tabular}{c} 
            \includegraphics[width=0.21\textwidth]{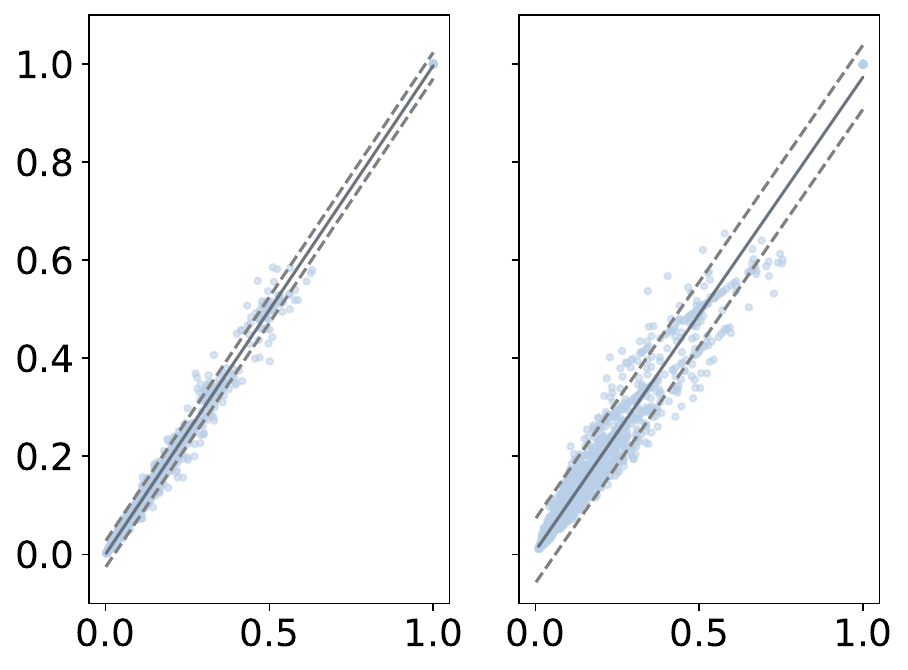} \\ 
            (4.46/26.08)
        \end{tabular}
        \label{fig:attn_sa2}
            }
              \hfil
        \subfloat[$\alpha^{(4)} = f(e^{(4)})$]{
        \begin{tabular}{c} 
            \includegraphics[width=0.21\textwidth]{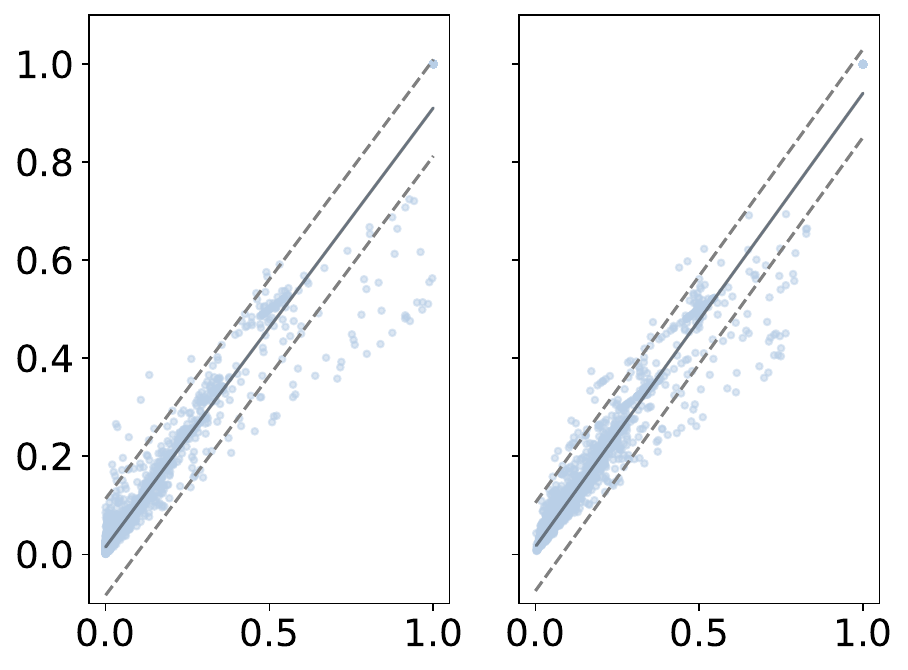} \\ 
            ($\inf$/50.61)
        \end{tabular}
        \label{fig:attn_sa3}
            }
              \hfil
              
        \subfloat[$\alpha^{(1)} = f(c)$]{
        \begin{tabular}{c} 
            \includegraphics[width=0.21\textwidth]{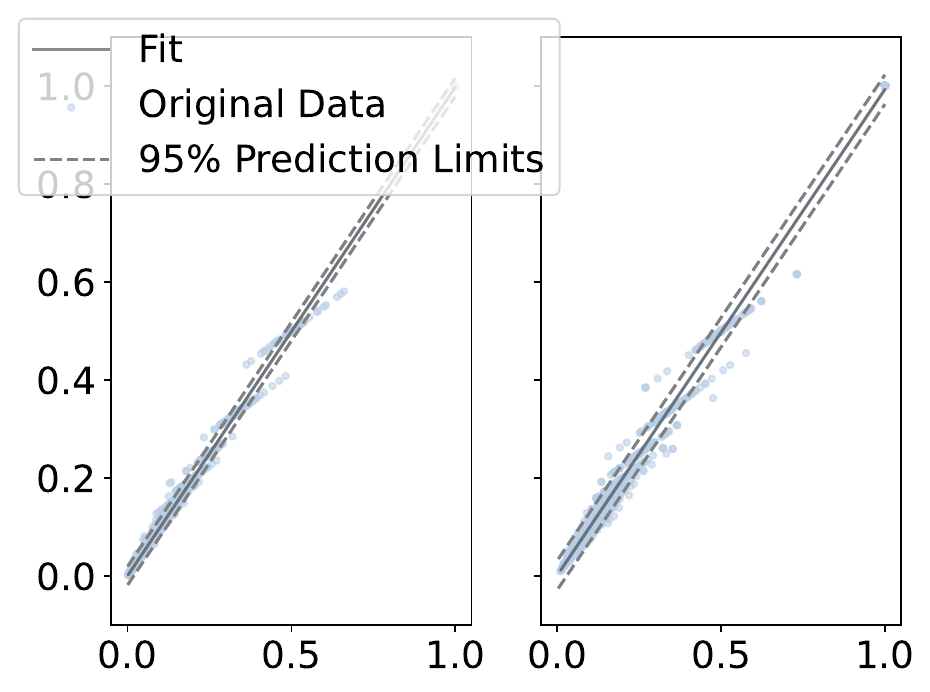} \\ 
            (2.21/5.26)
        \end{tabular}
        \label{fig:attn_sc0}
            }
            \hfil
        \subfloat[$\alpha^{(2)} = f(c)$]{
        \begin{tabular}{c} 
            \includegraphics[width=0.21\textwidth]{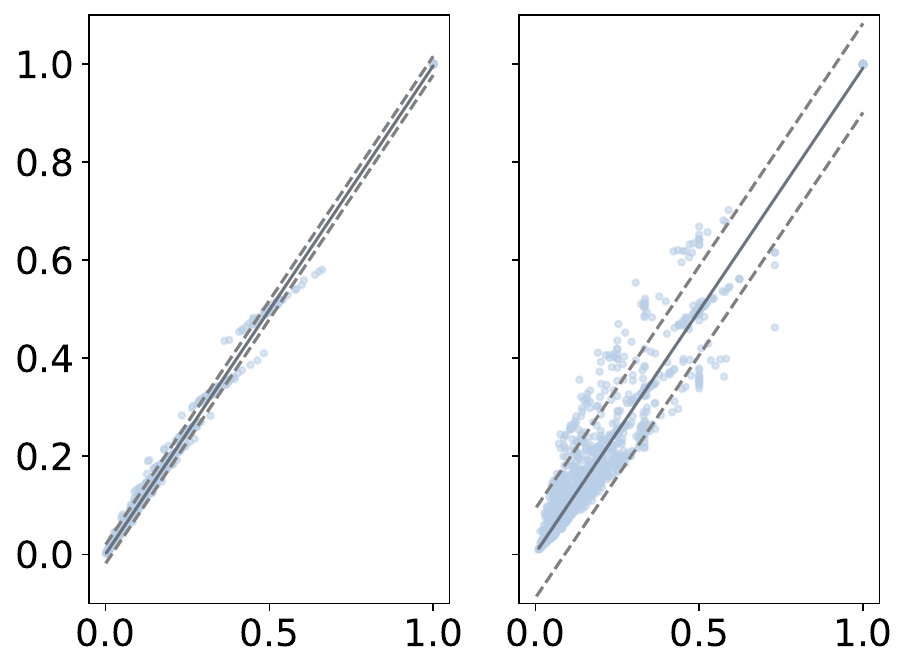} \\ 
            (2.32/49.07)
        \end{tabular}
        \label{fig:attn_sc1}
            }
            \hfil
        \subfloat[$\alpha^{(3)} = f(c)$]{
        \begin{tabular}{c} 
            \includegraphics[width=0.21\textwidth]{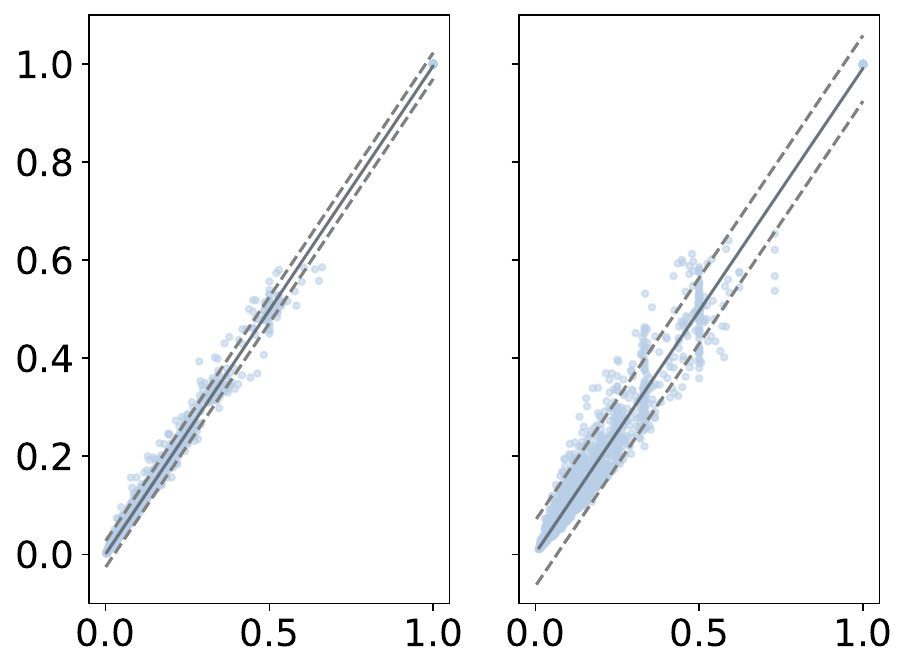} \\ 
            (4.98/27.61)
        \end{tabular}
        \label{fig:attn_sc2}
            }
              \hfil
        \subfloat[$\alpha^{(4)} = f(c)$]{
        \begin{tabular}{c} 
            \includegraphics[width=0.21\textwidth]{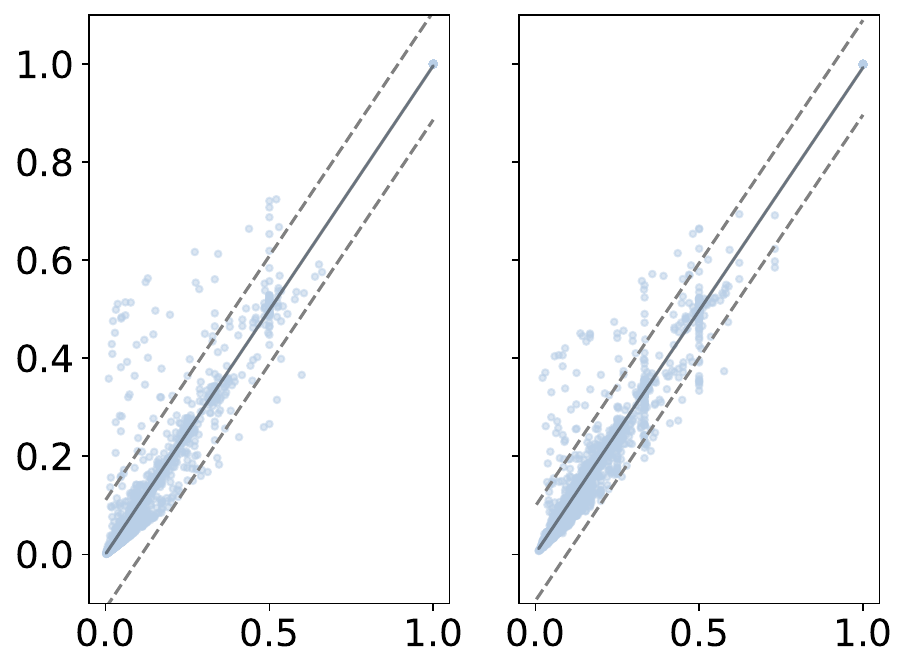} \\ 
            (110.59/69.58)
        \end{tabular}
        \label{fig:attn_sc3}
            }
              \hfil
      \caption{The outcomes of the linear regression analysis concerning the "cites"/"is cited by" scores within the field of computer science. Below the respective sub-figures, we depict the mean Kullback-Leibler (KL) divergences between the target and source distributions, denoted in units of $10^{-4}$.}
    \label{fig:attn}
    \end{figure*}
    
    To effectively visualize the connections between distributions, we gather all scores after applying softmax based on the target node. These scores are flattened to observe the overarching trend within the latest year. The original attention scores $e$ are averaged across all heads for ease of interpretation. Notably, we employ reversed values via $1 - c$ for the negative effect in the co-citing relationship, rendering them positive to reflect the trend.
    In Figure \ref{fig:attn}, robustly positive relationships are observed among the sum distribution $\alpha$, original attention distribution $e$, and the distribution of co-cited/citing strengths $c$. It is reasonable to note that the sum distribution closely resembles either of the sourced distributions. Compared to the co-cited/citing distribution, the sum distribution demonstrates greater similarity to the original attention distribution. Intriguingly, within the initial layers of the model, minimal disparities are evident between the original attention scores and the co-cited/citing strengths. This phenomenon is attributed to both the averaging across all heads and the model's adaptation to the co-cited/citing information during the learning process. Moreover, while the overall trend is evident, a finer examination of the samples reveals that for instances with varying co-cited/citing strengths, their original attention scores tend to be equal, resulting in the displayed horizontal and vertical distribution evident in Figures \ref{fig:attn_ac0} and \ref{fig:attn_sa0}. This equal might arise due to the initial state of the attention mechanism in the GAT.
    However, as the model progresses in-depth, noticeable disparities emerge between these distributions in both "cites" and "is cited by" edges. In the upper layers, attention scores tend to concentrate on a select few nodes, while in the lower layers, they disperse attention across a wider array of nodes. 
    Furthermore, co-cited/citing strengths represent 2-hop information conveyed through intermediate nodes, resembling lower-hop information within GNNs.
    Therefore, continuously incorporating the co-cited/citing distribution across all layers reinforces signals from pivotal nodes across a broader spectrum. It can alleviate the over-smoothing issues by preventing the dominance of a select few nodes within aggregated information.
    This approach serves to explain why a layer count of 4 emerges as optimal—it encapsulates higher-order information beyond the 2-hop range, allowing the co-cited/citing information to complement the original attention scores. This collaboration achieves multi-hop information integration, mitigating both forgetting and overemphasis within the learning process.

\subsection{Disentanglement Correlation Analysis}
    \begin{table*}[htbp]
      \centering
      \setstretch{0.8}
      \caption{Results of Disentanglement Correlation in Computer Science}
    \begin{tabular}{l|rrr|rrr|rrr}
    \toprule
    CS    & \multicolumn{1}{c}{dif} & \multicolumn{1}{c}{con} & \multicolumn{1}{c|}{contri} & \multicolumn{1}{c}{dif (\%)} & \multicolumn{1}{c}{con (\%)} & \multicolumn{1}{c|}{contri (\%)} & \multicolumn{1}{c}{citations} & \multicolumn{1}{c}{final} & \multicolumn{1}{c}{real} \\
    \midrule
    dif   & \multicolumn{1}{c}{--}  & 0.9579  & 0.9990  & 0.0389  & -0.3237  & 0.4556  & 0.3209  & 0.9971  & 0.8276  \\
    con   & 0.9579  & \multicolumn{1}{c}{--}  & 0.9620  & -0.1176  & -0.1544  & 0.3495  & 0.3575  & 0.9765  & 0.8091  \\
    contri & 0.9990  & 0.9620  & \multicolumn{1}{c|}{--}  & 0.0276  & -0.3236  & 0.4665  & 0.3239  & 0.9981  & 0.8287  \\
    \midrule
    dif (\%) & 0.0389  & -0.1176  & 0.0276  & \multicolumn{1}{c}{--}  & -0.7657  & 0.1967  & -0.0364  & -0.0022  & -0.0003  \\
    con (\%) & -0.3237  & -0.1544  & -0.3236  & -0.7657  & \multicolumn{1}{c}{--}  & -0.7812  & -0.0056  & -0.2861  & -0.2477  \\
    contri (\%) & 0.4556  & 0.3495  & 0.4665  & 0.1967  & -0.7812  & \multicolumn{1}{c|}{--}  & 0.0439  & 0.4382  & 0.3779  \\
    \midrule
    citations & 0.3209  & 0.3575  & 0.3239  & -0.0364  & -0.0056  & 0.0439  & \multicolumn{1}{c}{--}  & 0.3331  & 0.2982  \\
    final & 0.9971  & 0.9765  & 0.9981  & -0.0022  & -0.2861  & 0.4382  & 0.3331  & \multicolumn{1}{c}{--}  & 0.8298  \\
    real  & 0.8276  & 0.8091  & 0.8287  & -0.0003  & -0.2477  & 0.3779  & 0.2982  & 0.8298  & \multicolumn{1}{c}{--}  \\
    \midrule
    \midrule
    CHM   & \multicolumn{1}{c}{dif} & \multicolumn{1}{c}{con} & \multicolumn{1}{c|}{contri} & \multicolumn{1}{c}{dif (\%)} & \multicolumn{1}{c}{con (\%)} & \multicolumn{1}{c|}{contri (\%)} & \multicolumn{1}{c}{citations} & \multicolumn{1}{c}{final} & \multicolumn{1}{c}{real} \\
    \midrule
    dif   & \multicolumn{1}{c}{--}  & 0.8441  & 0.9932  & 0.6134  & -0.3982  & -0.0478  & 0.1844  & 0.9828  & 0.7834  \\
    con   & 0.8441  & \multicolumn{1}{c}{--}  & 0.8723  & 0.2398  & -0.0062  & -0.3362  & 0.2581  & 0.9267  & 0.7317  \\
    contri & 0.9932  & 0.8723  & \multicolumn{1}{c|}{--}  & 0.5724  & -0.3795  & -0.0275  & 0.2035  & 0.9911  & 0.7912  \\
    \midrule
    dif (\%) & 0.6134  & 0.2398  & 0.5724  & \multicolumn{1}{c}{--}  & -0.9104  & 0.4773  & -0.0080  & 0.5075  & 0.4006  \\
    con (\%) & -0.3982  & -0.0062  & -0.3795  & -0.9104  & \multicolumn{1}{c}{--}  & -0.7981  & 0.0477  & -0.2885  & -0.2331  \\
    contri (\%) & -0.0478  & -0.3362  & -0.0275  & 0.4773  & -0.7981  & \multicolumn{1}{c|}{--}  & -0.0897  & -0.1263  & -0.0883  \\
    \midrule
    citations & 0.1844  & 0.2581  & 0.2035  & -0.0080  & 0.0477  & -0.0897  & \multicolumn{1}{c}{--}  & 0.2189  & 0.2022  \\
    final & 0.9828  & 0.9267  & 0.9911  & 0.5075  & -0.2885  & -0.1263  & 0.2189  & \multicolumn{1}{c}{--}  & 0.7955  \\
    real  & 0.7834  & 0.7317  & 0.7912  & 0.4006  & -0.2331  & -0.0883  & 0.2022  & 0.7955  & \multicolumn{1}{c}{--}  \\
    \midrule
    \midrule
    PSY   & \multicolumn{1}{c}{dif} & \multicolumn{1}{c}{con} & \multicolumn{1}{c|}{contri} & \multicolumn{1}{c}{dif (\%)} & \multicolumn{1}{c}{con (\%)} & \multicolumn{1}{c|}{contri (\%)} & \multicolumn{1}{c}{citations} & \multicolumn{1}{c}{final} & \multicolumn{1}{c}{real} \\
    \midrule
    dif   & \multicolumn{1}{c}{--}  & 0.8642  & 0.9972  & 0.5355  & -0.5871  & 0.6302  & 0.3556  & 0.9903  & 0.8196  \\
    con   & 0.8642  & \multicolumn{1}{c}{--}  & 0.8740  & 0.2039  & -0.2505  & 0.2997  & 0.4998  & 0.9244  & 0.7598  \\
    contri & 0.9972  & 0.8740  & \multicolumn{1}{c|}{--}  & 0.5014  & -0.5609  & 0.6149  & 0.3572  & 0.9928  & 0.8199  \\
    \midrule
    dif (\%) & 0.5355  & 0.2039  & 0.5014  & \multicolumn{1}{c}{--}  & -0.9869  & 0.9337  & -0.0134  & 0.4522  & 0.3917  \\
    con (\%) & -0.5871  & -0.2505  & -0.5609  & -0.9869  & \multicolumn{1}{c}{--}  & -0.9792  & 0.0056  & -0.5070  & -0.4328  \\
    contri (\%) & 0.6302  & 0.2997  & 0.6149  & 0.9337  & -0.9792  & \multicolumn{1}{c|}{--}  & 0.0043  & 0.5570  & 0.4683  \\
    \midrule
    citations & 0.3556  & 0.4998  & 0.3572  & -0.0134  & 0.0056  & 0.0043  & \multicolumn{1}{c}{--}  & 0.4017  & 0.3416  \\
    final & 0.9903  & 0.9244  & 0.9928  & 0.4522  & -0.5070  & 0.5570  & 0.4017  & \multicolumn{1}{c}{--}  & 0.8256  \\
    real  & 0.8196  & 0.7598  & 0.8199  & 0.3917  & -0.4328  & 0.4683  & 0.3416  & 0.8256  & \multicolumn{1}{c}{--}  \\
    \bottomrule
    \end{tabular}%
      \label{tab:corr}%
    \end{table*}%
    
    In Table \ref{tab:corr}, we illustrate the correlations among all disentangled perspective values/proportions and the predicted/real values, focusing on samples with all positively predicted values across all datasets.
    
    Taking the correlation analysis of Computer Science as an example, "conformity" stands out among the values of all three perspectives. Its correlation with the predicted/real values shows a significant gap (-0.02) compared to other perspectives. Despite this, conformity emerges as the most closely associated perspective with accumulated citations, implying its significance in representing information about a paper's reputation.
    The predicted increment appears as the most influential predictor of the real increment. Its aggregated value outperforms individual perspectives, highlighting the effectiveness of the proposed disentangled learning approach. Moreover, the real increment demonstrates minimal correlation with the paper's accumulated citations. This revelation underscores two key points: (1) Citation increment proves to be an effective metric in portraying a paper's potential impact, with accumulated citation accounting for only a fraction of the future increment. (2) The DPPDCC model adeptly captures the trend in a paper's citation increment, validating its substantial potential and reliability.
    In our calculations of the proportions of all three perspectives within predicted values, we find that conformity diffusion is the only perspective with numerous negative correlations. This suggests that in the field of Computer Science, papers characterized solely by accumulated citations tend to attract fewer new citations. Additionally, it shows an almost negative correlation with the proportion of contribution (-0.78), indicating their mutual exclusivity. Consequently, conformity and contribution have learned entirely distinct facets of citation increments.
    Overall, the contribution perspective is the most pivotal factor for prediction, showcasing the highest correlations in both values and proportions. However, the correlations remain a flexible threshold, where other perspectives also help determine the predicted increments. As a complement to the numerical citation increment, this approach offers a more objective evaluation, potentially aiding in identifying valuable papers by encompassing diverse and complementary information.

    In summary, contribution is the most significant perspective across all datasets. Its disentangled value has the largest correlation with both the predicted and real citation increments, as does its proportion in Computer Science and Psychology. Moreover, diffusion and conformity focus on different aspects of the popularity factor. Accumulated citations consistently explain the trend of conformity values, while diffusion complements the structural influence with more flexibility regarding graph features.

\subsection{Complete Disentanglement Composition}

    \begin{figure*}[]
      \centering
            \subfloat[time trend (CS)]{\includegraphics[height=0.14\textheight]{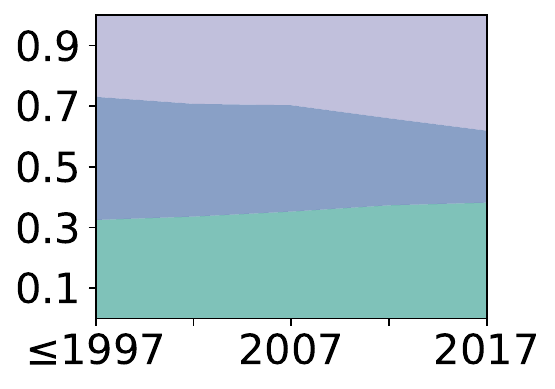}%
    		\label{fig:vis_time_cs}}
            \hfil
    	\subfloat[group detail (CS)]{\includegraphics[height=0.14\textheight]{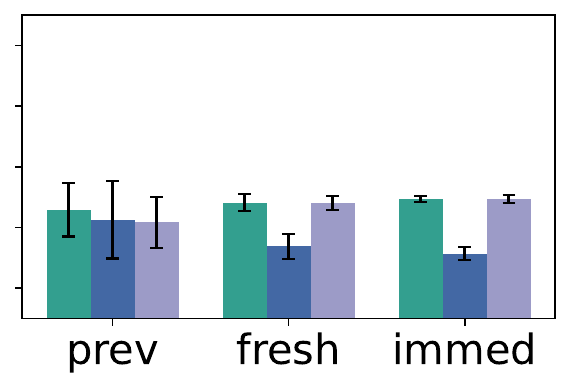}%
    		\label{fig:vis_group_cs}}
            \hfil
    	\subfloat[value bin (CS)]{\includegraphics[height=0.14\textheight]{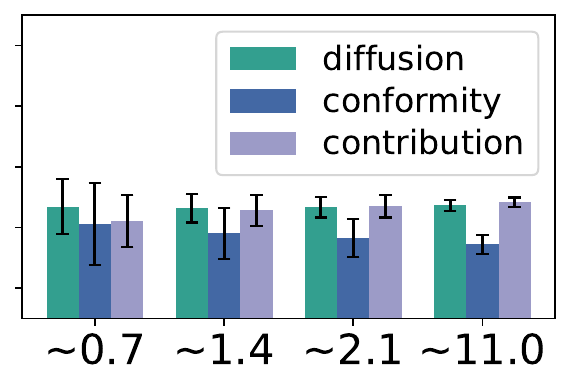}%
    		\label{fig:vis_vbin_cs}}
            \hfil
            \subfloat[time trend (CHM)]{\includegraphics[height=0.14\textheight]{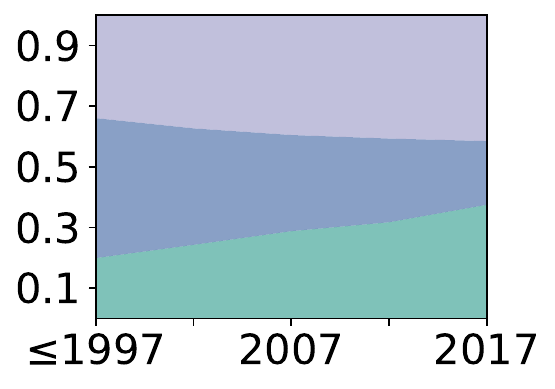}%
    		\label{fig:vis_time_chm}}
            \hfil
    	\subfloat[group detail (CHM)]{\includegraphics[height=0.14\textheight]{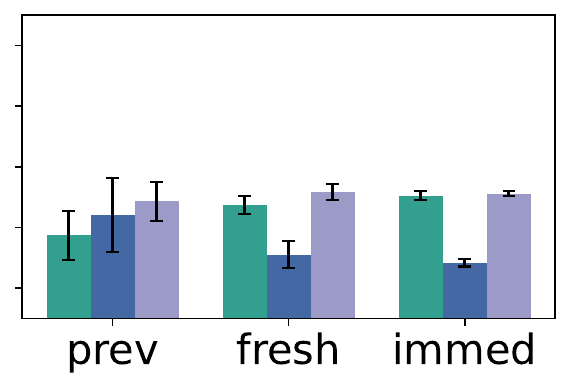}%
    		\label{fig:vis_group_chm}}
            \hfil
    	\subfloat[value bin (CHM)]{\includegraphics[height=0.14\textheight]{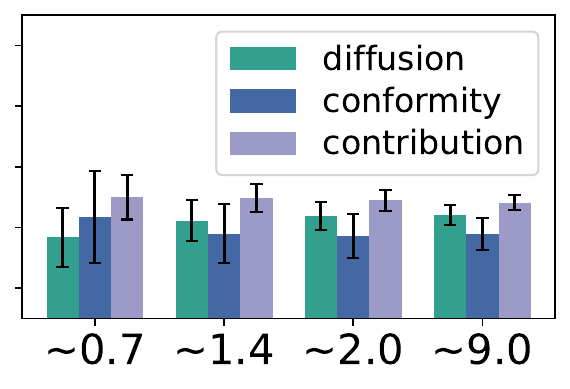}%
    		\label{fig:vis_vbin_chm}}
            \hfil
            \subfloat[time trend (PSY)]{\includegraphics[height=0.14\textheight]{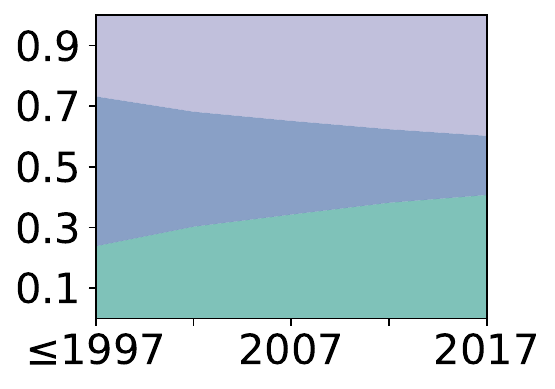}%
    		\label{fig:vis_time_psy}}
            \hfil
    	\subfloat[group detail (PSY)]{\includegraphics[height=0.14\textheight]{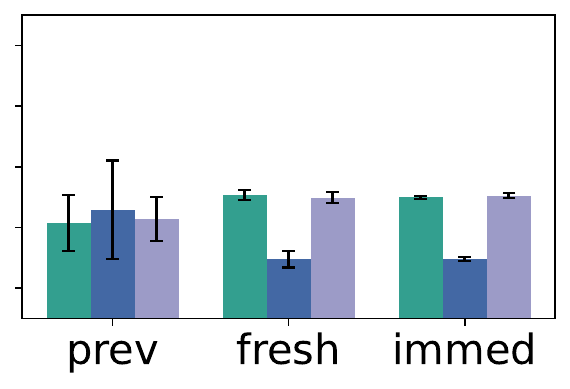}%
    		\label{fig:vis_group_psy}}
            \hfil
    	\subfloat[value bin (PSY)]{\includegraphics[height=0.14\textheight]{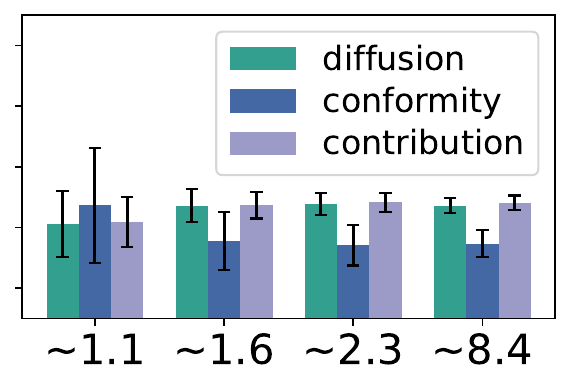}%
    		\label{fig:vis_vbin_psy}}
            \hfil
      \caption{Visualization of disentangled value proportions. (a)/(d)/(g) displays the trend that evolved with the publication time. (b)/(e)/(h) demonstrates the detailed composition of papers categorized into previous, fresh, and immediate ones. (c)/(f)/(i) is binning the samples based on the predicted values.}
    \label{fig:vis_comp}
    \end{figure*}

    Figure \ref{fig:vis_comp} illustrates the disentangled compositions across all datasets. We observe common trends, such as the similar pattern of conformity and the dominance of contribution. Additionally, details such as the respective slopes and nuanced composition for various categories vary according to the intrinsic characteristics of the corresponding fields.

\section{Case Study}
    \begin{table*}[htbp]
      \centering
      \setstretch{0.8}
      \caption{Detailed meta-data, disentangled values/proportions, and potential impact of the selected cases.}
\begin{tabular}{r|rrr|rrr|rrr|rr}
    \toprule
    \multicolumn{1}{c|}{\multirow{2}[4]{*}{id}} & \multicolumn{3}{c|}{meta} & \multicolumn{3}{c|}{disentangled values} & \multicolumn{3}{c|}{disentangled proportions} & \multicolumn{2}{c}{potential impact} \\
\cmidrule{2-12}          & \multicolumn{1}{c}{pub time} & \multicolumn{1}{c}{refs} & \multicolumn{1}{c|}{citations} & \multicolumn{1}{c}{dif} & \multicolumn{1}{c}{con} & \multicolumn{1}{c|}{contri} & \multicolumn{1}{c}{dif} & \multicolumn{1}{c}{con} & \multicolumn{1}{c|}{contri} & \multicolumn{1}{c}{pred} & \multicolumn{1}{c}{real} \\
    \midrule
    981400 & 2016  & 28    & 7 (3/2) & 1.2175  & 0.7428  & 1.1862  & 38.69\% & 23.61\% & 37.70\% & 3.1466  & 3.0445  \\
    47429 & 2009  & 12    & 51 (15/3) & 1.1432  & 0.7895  & 1.1993  & 36.50\% & 25.21\% & 38.29\% & 3.1320  & 3.1355  \\
    88781 & 2014  & 37    & 5 (1/0) & 1.1984  & 0.7247  & 1.2307  & 38.00\% & 22.98\% & 39.02\% & 3.1538  & 4.7449  \\
    \bottomrule
    \end{tabular}%
      \label{tab:case_study}%
    \end{table*}%

    \begin{figure*}[]
      \centering
        \centering
        \subfloat[981400-$c$]{\includegraphics[width=0.33\textwidth]{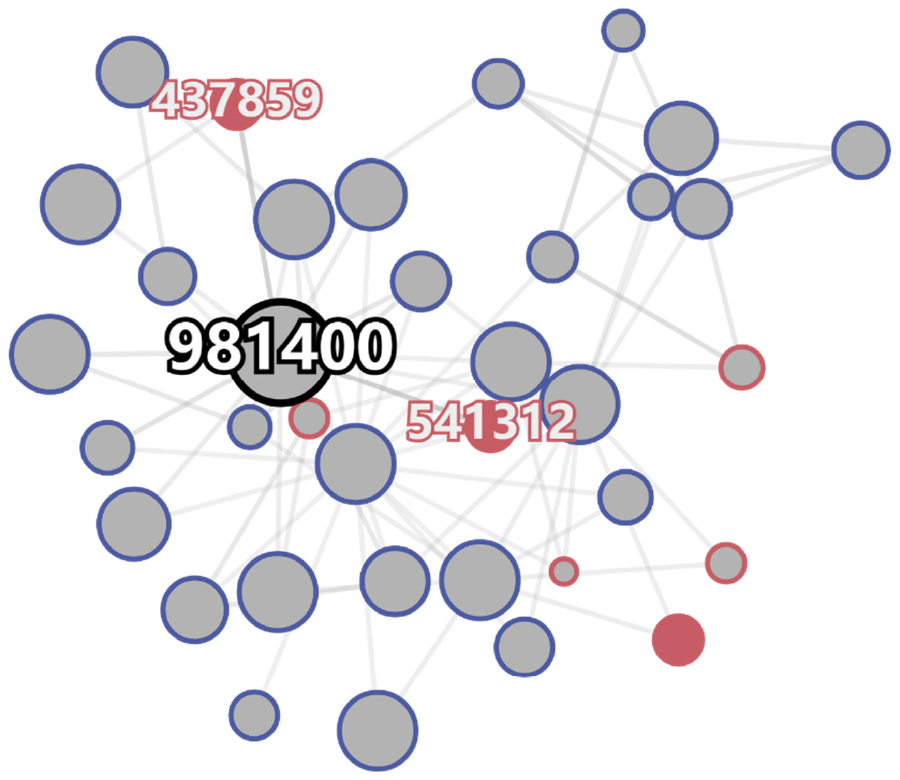}%
            \label{fig:dif_c}}
            \hfil
        \subfloat[981400-$e$]{\includegraphics[width=0.33\textwidth]{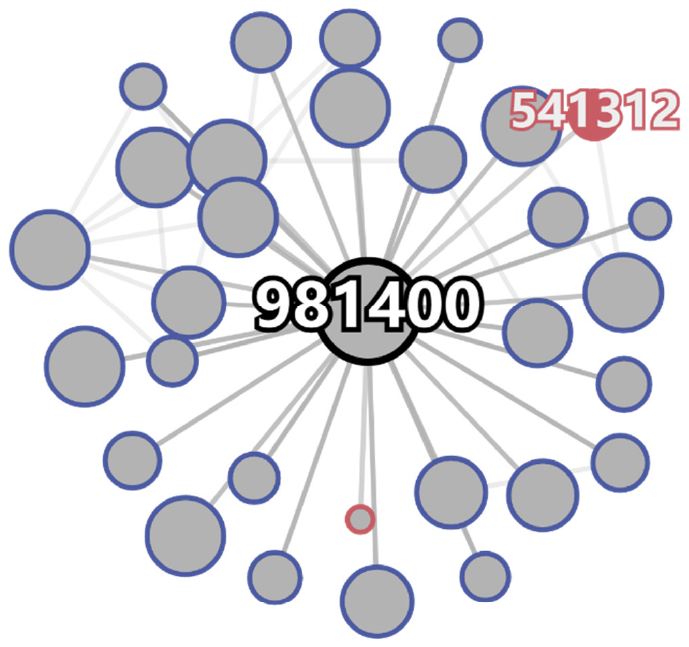}%
            \label{fig:dif_a}}
            \hfil
        \subfloat[981400-$\alpha$]{\includegraphics[width=0.33\textwidth]{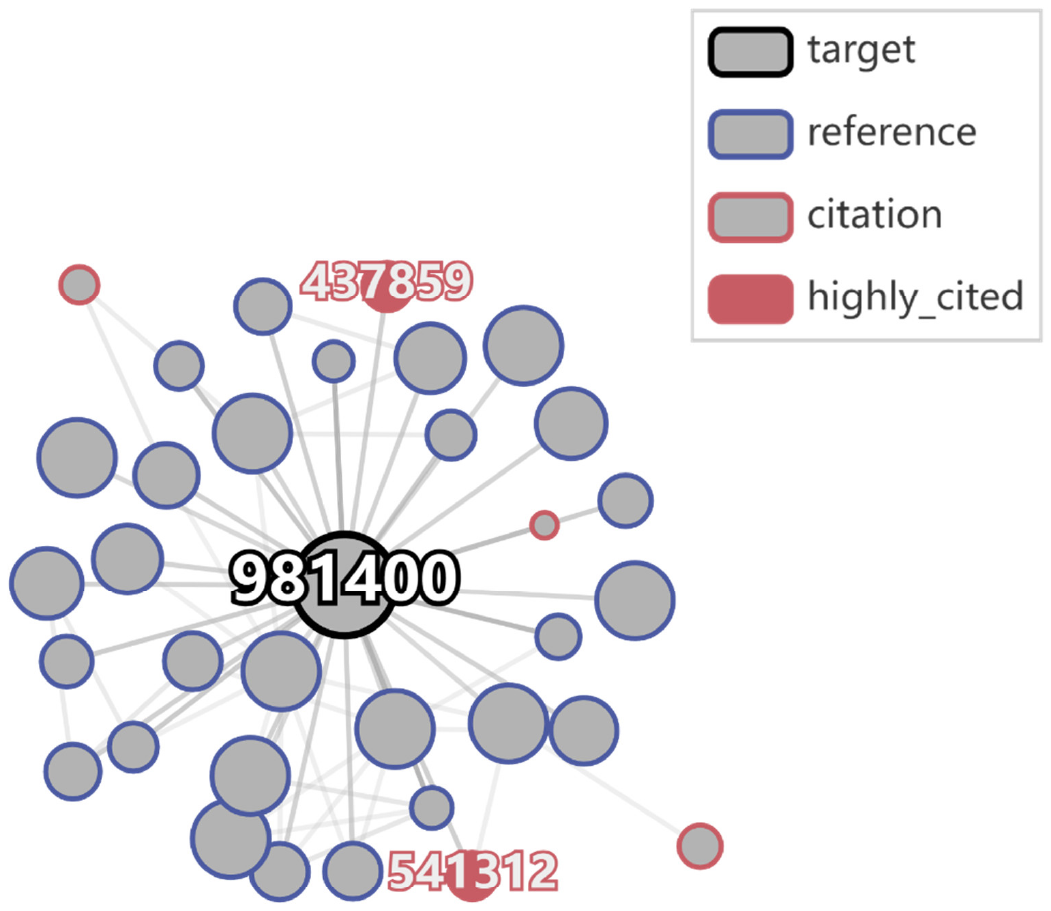}%
            \label{fig:dif_s}}
              \hfil
              
        \subfloat[47429-$c$]{\includegraphics[width=0.33\textwidth]{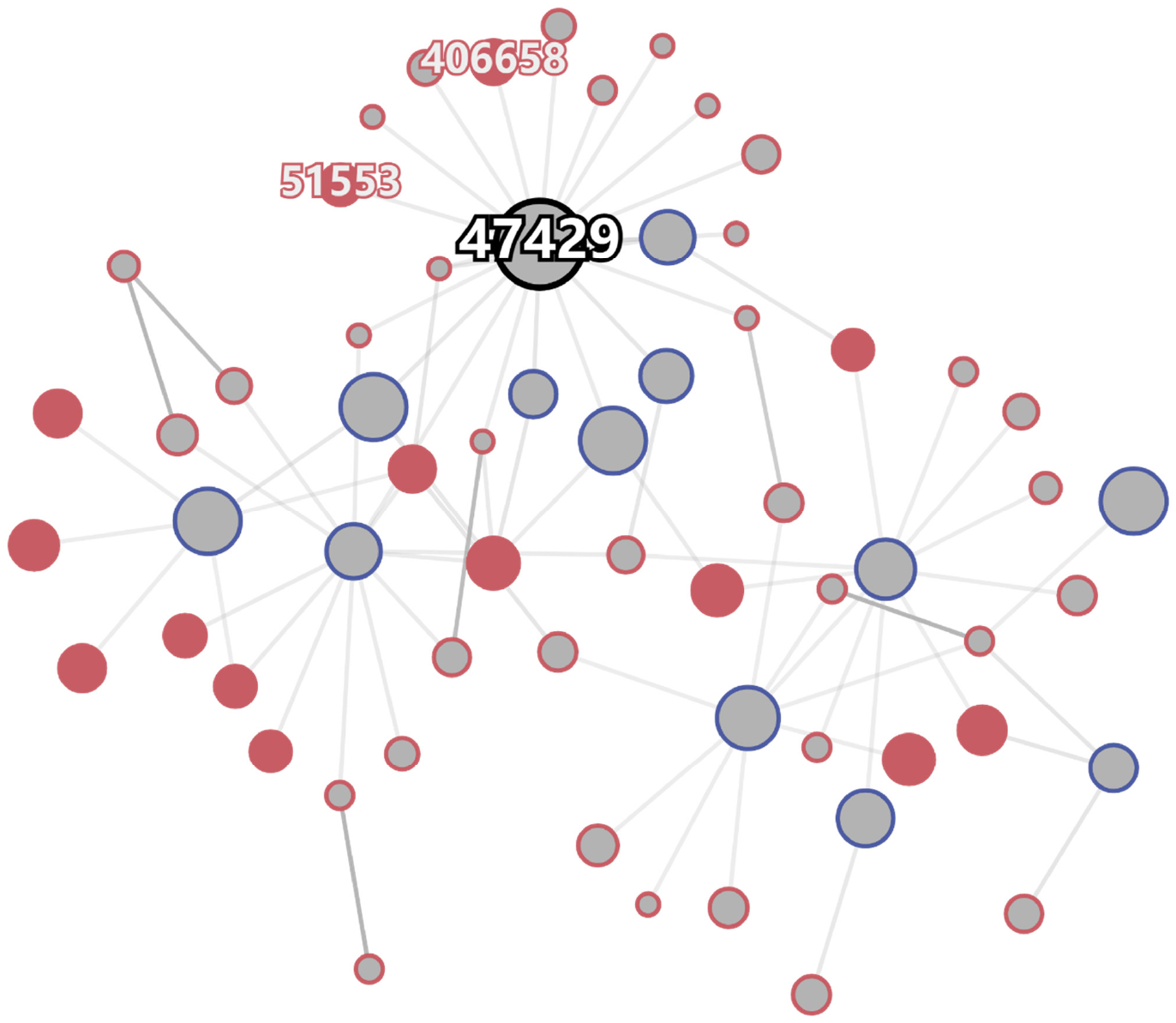}%
            \label{fig:con_c}}
            \hfil
        \subfloat[47429-$e$]{\includegraphics[width=0.33\textwidth]{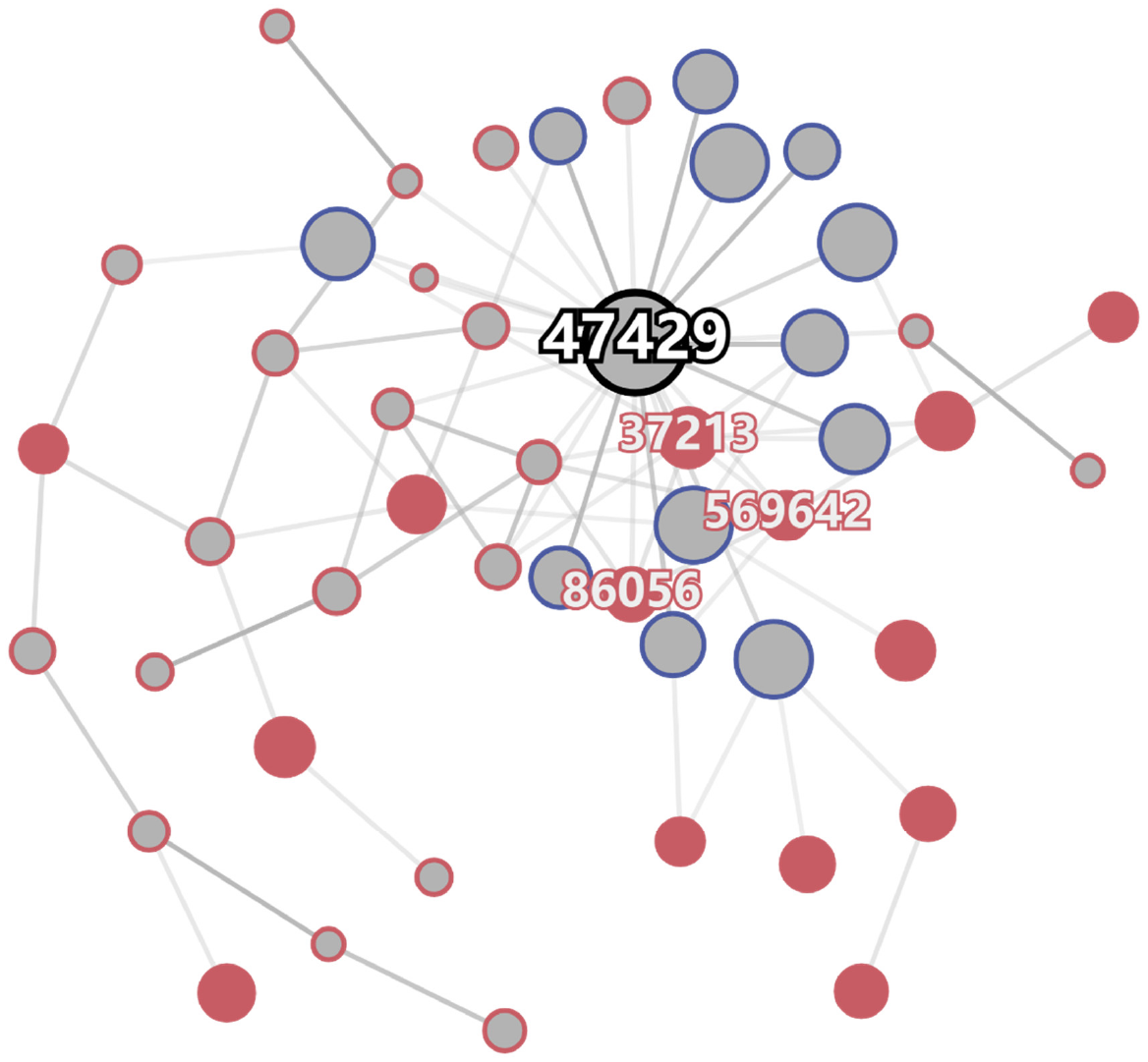}%
            \label{fig:con_a}}
            \hfil
        \subfloat[47429-$\alpha$]{\includegraphics[width=0.33\textwidth]{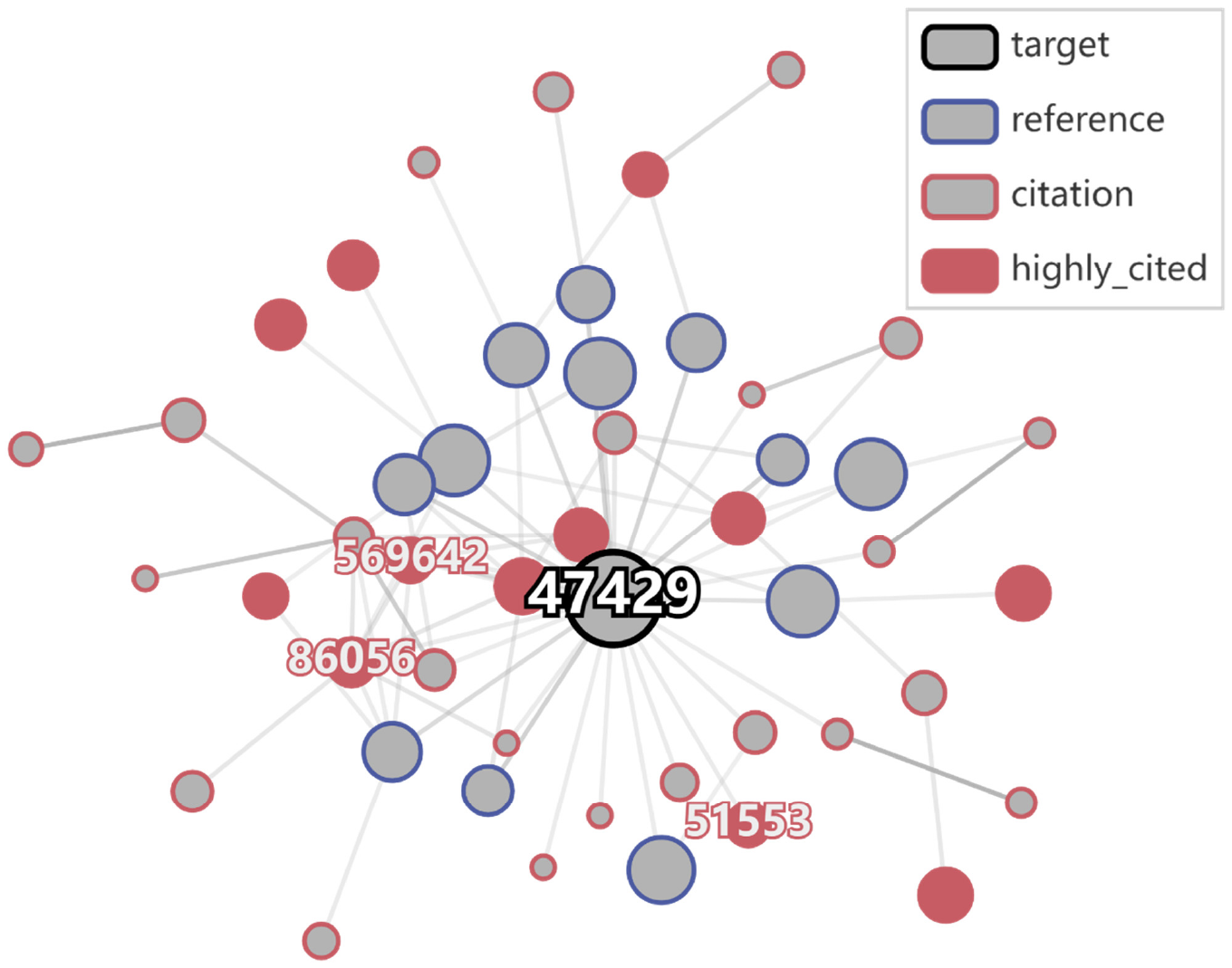}%
            \label{fig:con_s}}
              \hfil
              
        \subfloat[88781-$c$]{\includegraphics[width=0.33\textwidth]{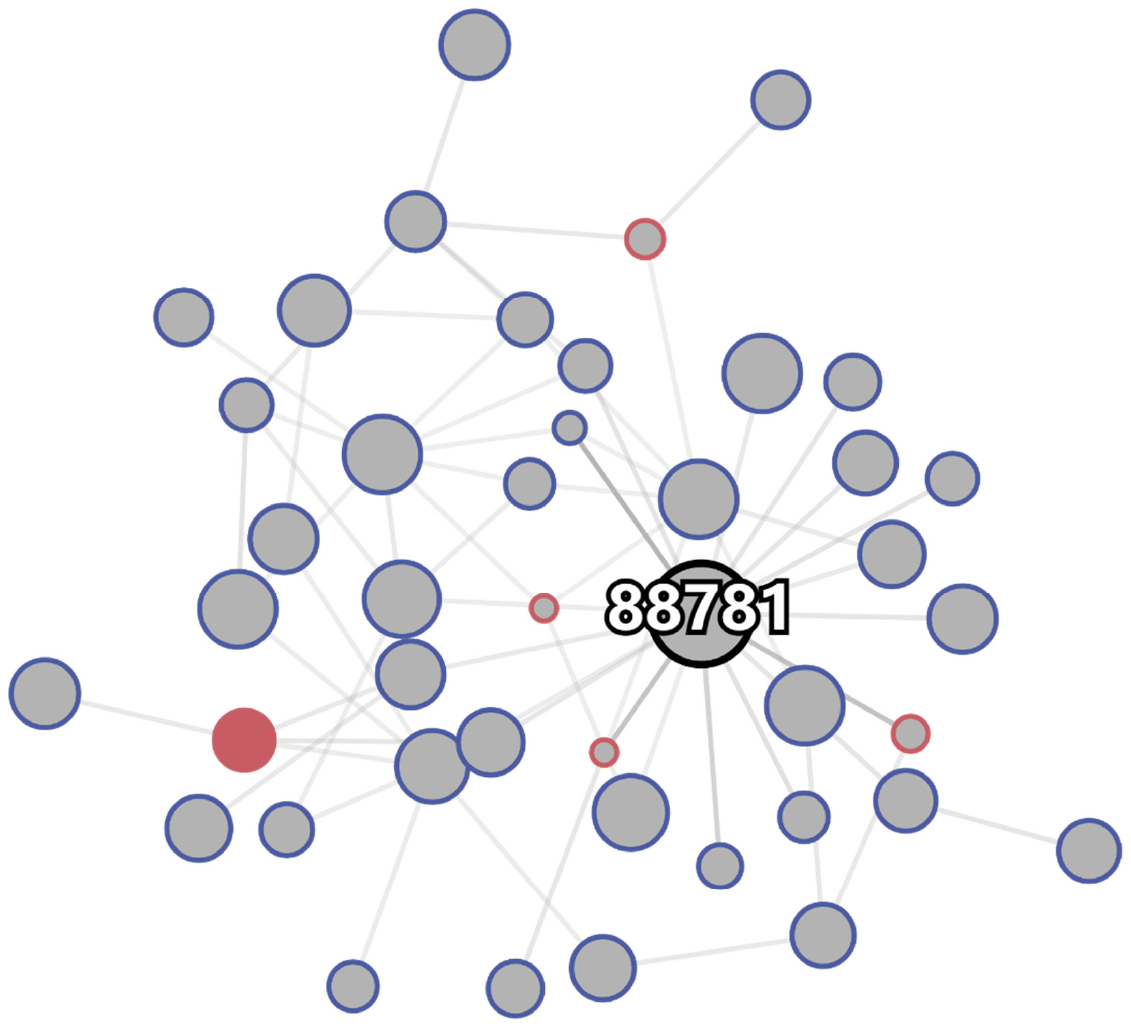}%
            \label{fig:contri_c}}
            \hfil
        \subfloat[88781-$e$]{\includegraphics[width=0.33\textwidth]{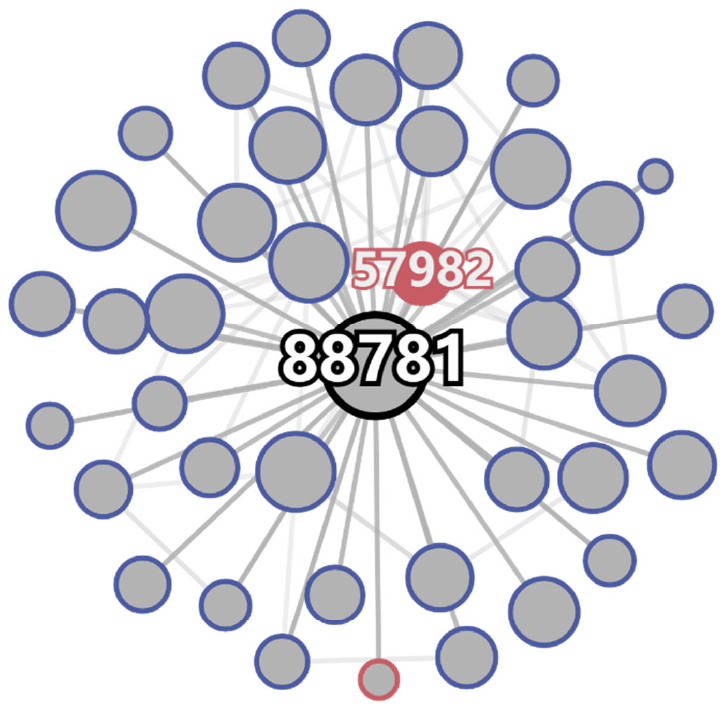}%
            \label{fig:contri_a}}
            \hfil
        \subfloat[88781-$\alpha$]{\includegraphics[width=0.33\textwidth]{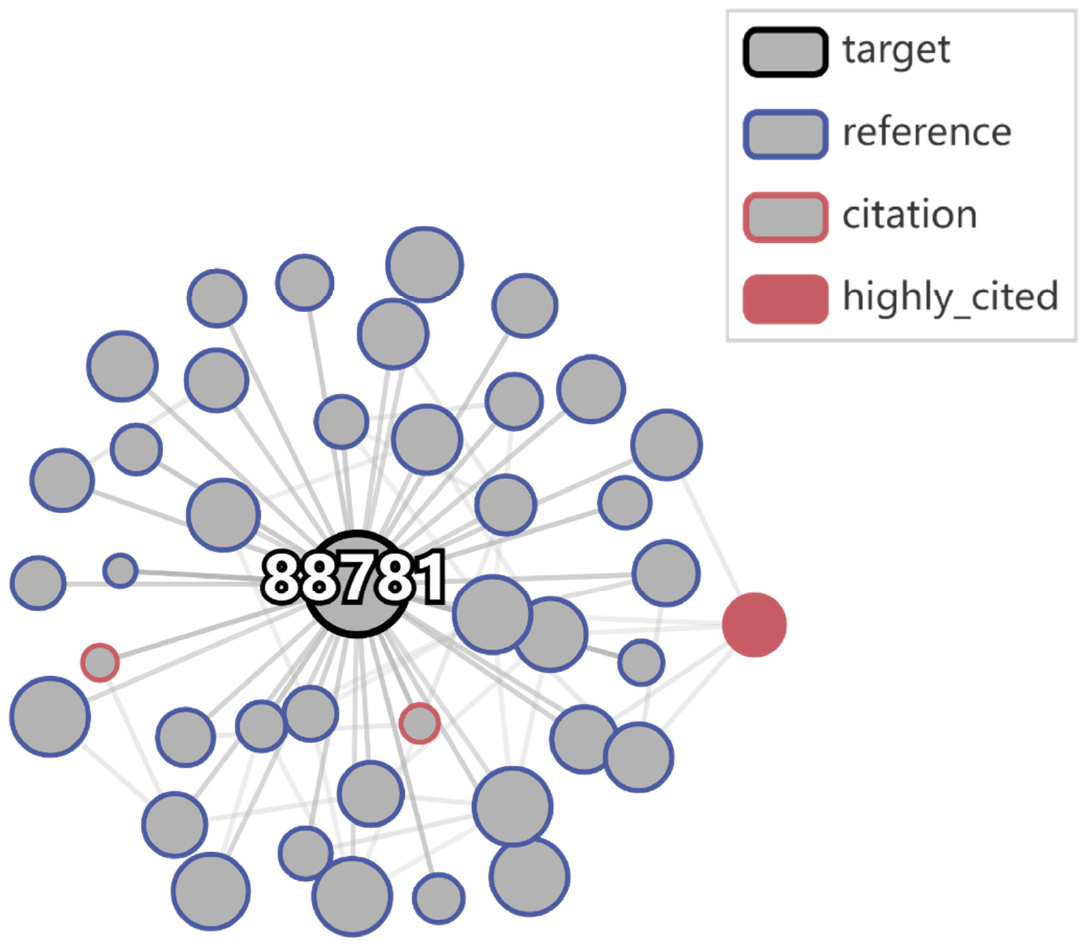}%
            \label{fig:contri_s}}
              \hfil
      \caption{Weighted \textit{paper} "cites" subgraphs of selected cases organized by co-cited strengths $c$, average original attention scores $e$, and sum scores $\alpha$ in Computer Science. We filter the edges within 2-hop graphs by retaining only the top 30\% accumulated weights of each node and preserve the direct citation or reference nodes of the target paper to attain these sparser graphs. We only label the target paper and its directly linked highly-cited citations.}
    \label{fig:case_study}
    \end{figure*}

    To delve into the intricate workings of our proposed model in addressing \textbf{RQ4}, we select three representative papers for each perspective. These papers should be co-cited or citing papers with similar topics, reflecting comparable predicted and real potential impact. Our approach involves contrasting their metadata, disentangled values, perspective proportions, and predicted/real potential impact. In this section, we visualize graph structures organized by co-cited strengths $c$, average original attention scores $e$, and sum scores $\alpha$ within "cites" edge subgraphs of the selected papers. To maintain clarity, we focus on graphs extracted from the latest year of the snapshots and average scores across all layers of DPPDCC. We then filter the edges in 2-hop graphs based on the top 30\% accumulated weights, resulting in sparser graphs. To create target-centric connected graphs, we retain only the direct citations and references of the target paper. Highly-cited papers, representing papers that provide the top 80\% of citations within their respective fields of the whole datasets, are highlighted by filling in red within the graphs. Our labeling strategy encompasses the target paper and its directly connected highly-cited citations.
    
    As outlined in Table \ref{tab:case_study}, papers 981400, 47429, and 88781, respectively, embody the diffusion, conformity, and contribution perspectives. Despite their near-identical predicted increments, their disentangled value compositions differ from one another. Notably, 981400 and 88781 are recent papers not encountered during the training phase, while 47429 is an earlier publication, each with varying years since publication and distinct accumulated citations. These papers exhibit perspective proportions aligning with all trends discussed in Section \ref{subsec:disen_visual}.
    As depicted in Figure \ref{fig:case_study}, $c$ subgraphs manifest as the densest, while $e$ and $\alpha$ subgraphs appear considerably sparser, featuring fewer interconnected nodes. In the higher layers of DPPDCC, the model tends to concentrate on specific nodes, resulting in sparsity in these graphs. The introduced co-cited strengths $c$ act as a complementary factor to the original attention scores $e$, facilitating the identification of significant citations from highly-cited papers. Notably, $c$ consistently integrates significant highly-cited papers that were overlooked, replacing the original set with more valuable ones in $e$ subgraphs.
    For instance, Figure \ref{fig:dif_a} considers only one highly-cited paper (541312). However, in Figure \ref{fig:dif_s}, one additional highly-cited paper (437859) - initially presented in Figure \ref{fig:dif_c} is taken into account. Similarly, although Figure \ref{fig:dif_s} and \ref{fig:dif_a} share the same number of highly-cited papers (3), the co-cited subgraph $c$ replaces one of them (37213 $\rightarrow$ 51553).
    
    For paper 981400, within the test set, it stands as a recent publication with limited citations due to its relatively short existence. Although it has fewer accumulated citations than paper 47429, 981400 has attracted significant attention from two highly-cited papers, which is noteworthy given its total of seven citations. Consequently, its average influential citation (28.6\%) exceeds that of 47429 (5.9\%), resulting in a higher proportion of the diffusion perspective in its incremental source. This positions it as a crucial element in the information diffusion process, contributing to its high diffusion values.
    On the other hand, paper 47429, the oldest paper in the test set with a publication span of nearly 8 years by 2017, stands out with the highest conformity value among all cases, significantly contributing to a larger proportion of its incremental impact. Despite its extensive citation count, its significance has diminished over time in the evolving context of the citation network, with citations tending to emphasize its established reputation rather than other factors.
    Paper 88781, also recent with limited citations, lacks both highly-cited and accumulated citations compared to papers 981400 or 47429. Consequently, it exhibits low values and proportions in both diffusion and conformity perspectives.
    However, its contribution perspective emerges as pivotal, showcasing the largest values and proportion. Leveraging its contribution, 88781 surpasses the other two papers in predicted/real potential impact. Interestingly, DPPDCC highlights intrinsic factors to its contribution beyond popularity-related factors highlighted in graph structures or accumulated citation counts, emphasizing DPPDCC's substantial potential for practical application and impact.

    Indeed, our proposed DPPDCC effectively disentangles citation increments from different factors for the target paper. This disentanglement, as evidenced by the values and proportions assigned to distinct perspectives, offers more comparable metrics for identifying valuable papers. In future endeavors, we aspire to expand this framework and evaluate its practical utility in uncovering novel insights within real-world scenarios. This continued exploration will allow us to delve further into discoveries and new applications facilitated by our disentanglement approach.

\section{Limitations}
    In our future endeavors, we intend to develop extensive datasets spanning diverse fields of study to simulate practical scenarios. This initiative is geared towards significantly enhancing the practicality and relevance of our approach. Additionally, we plan to incorporate additional factors that account for citation intent, thereby enabling a more precise evaluation of the genuine contributions made by academic papers. Furthermore, we will conduct a thorough analysis of the model's performance in identifying novel papers within specific domains. 
    Importantly, we'll validate these findings by comparing results with relevant studies, elucidating the practical implications of our proposed methodology.
\twocolumn

\end{document}